\documentclass[prb,twocolumn,showpacs,preprintnumbers,superscriptaddress]{revtex4-1}
\usepackage[usenames,dvipsnames]{color}
\usepackage{amssymb} 
\usepackage{times}
\usepackage{graphicx}
\usepackage{graphicx}
\usepackage{dcolumn}
\usepackage{datetime}
\usepackage{soul}
\usepackage{subfigure}
\usepackage[bookmarks, colorlinks=true, plainpages = false,citecolor = red, urlcolor = black, filecolor = blue]{hyperref}
\usepackage{multirow} 
\usepackage[english]{babel}
\usepackage{graphicx}
\usepackage{amsmath}

\usepackage[english]{babel}
\usepackage{graphicx}
\usepackage{amsmath}
 
\begin{document}
\title{Tight Binding Model of Mn$_{12}$ Single Molecule Magnets: Electronic and Magnetic Structure and Transport Properties}
 
\author{Fatemeh Rostamzadeh Renani} 
\affiliation{Department of Physics, Simon Fraser University, Burnaby, British Columbia, Canada V5A 1S6}

\author{George Kirczenow} 
\altaffiliation{Canadian Institute for Advanced Research, Nanoelectronics Program.}
\affiliation{Department of Physics, Simon Fraser
University, Burnaby, British Columbia, Canada V5A 1S6}

\date{\today}

\begin{abstract}\noindent
We describe and analyze  a tight-binding model of single molecule magnets (SMMs) that captures both the spin and spatial aspects of the SMM electronic structure. The model generalizes extended H\"{u}ckel theory to include the effects of spin polarization and spin-orbit coupling. For neutral and negatively charged Mn$_{12}$ SMMs with acetate or benzoate ligands the model yields the total SMM spin, the spins of the individual Mn ions, the magnetic easy axis orientation, the size of the magnetic anisotropy barrier and the size of the HOMO-LUMO gap consistent with experiment. For neutral molecules the predicted spins and spatial locations of the HOMO are consistent with the results of density functional calculations. For the total spin and location of the LUMO density functional theory-based calculations yield varied results while the present model yield results consistent with experiments on negatively charged molecules. 
For Mn$_{12}$ SMMs with thiolate- and methylsulphide-terminated benzoate ligands (Mn$_{12}$-Ph-Th) we find the HOMO to be located on the magnetic core of the molecule, but (unlike for the Mn$_{12}$ SMMs that have previously been studied theoretically), we predict the LUMO and near LUMO orbitals of  Mn$_{12}$-Ph-Th to be located on ligands. Therefore we predict that for these Mn$_{12}$ SMMs resonant and off-resonant coherent transport via near-LUMO orbitals, {\em not subject to Coulomb blockade}, should occur. We propose that this effect can be used to identify specific experimentally realized SMM transistors in which the easy axis and magnetic moment are approximately parallel to the direction of the current flow. We also predict effective spin filtering by these SMMs to occur at low bias whether the transport is mediated by the HOMO that is on the magnetic core of the SMM or by near LUMO orbitals located on the nominally non-magnetic ligands. 

\end{abstract}
%Electronic transport in nanoscale materials and structures, 73.63.-b,
%molecular magnets (magnetic materials), 75.50.Xx
%spin polarized transport devices, 85.75.-d
%Coulomb blockade, 73.23.Hk

\maketitle
% ----------------------------------------Section - 1-------------------------------------
\section{Introduction}
Single molecule magnets (SMMs) are magnetic molecules that contain transition metal atoms surrounded by organic ligands. \cite{SMM2008, SMMbookGatteschi, SMMChristou, SMMGatteschi}  In a SMM crystal, these ligands separate the magnetic core of each molecule from those of its neighbors making the inter-molecule magnetic exchange interactions negligible. Thus, the magnetic properties of a SMM crystal are due to the magnetic moments of the individual SMMs. The SMM's large spin and magnetic anisotropy barrier (MAB) hamper magnetization reversal below the SMM's blocking temperature. Mn$_{12}$, the first discovered SMM \cite{Lis}, contains four Mn$^{4+}$ ions (henceforth referred to as the ``inner Mn") surrounded by a nonplanar ring of eight Mn$^{3+}$ ions (the ``outer Mn") that interact antiferomagnetically with the inner Mn. The large ground state spin S=10~~\cite{S10} and large magnetic anisotropy barrier (up to $\sim 6.1$ meV along the easy axis)\cite{expt_MAB} lead to high blocking temperatures ($\sim 3.5$ K) \cite{MnLigandEffectZagaynova} and long relaxation times.\cite{expt_RelaxationTime} For these reasons the Mn$_{12}$ family has been the most studied among SMMs.

Electron and spin transport through individual molecules bridging a pair of electrodes, and through single molecule transistors that include a third ``gate" electrode, have been studied intensively for more than a decade.\cite{review2010} Because of their large magnetic anisotropy barriers and associated stable magnetic moments, single molecule magnets bring a new dimension to this field and also raise the possibility of molecular magnetic information storage. Therefore the transport properties of transistors based on individual SMMs are attracting considerable interest at present, both experimentally \cite{Heersche2006,moonHo2006, Zyazin2010} and theoretically.\cite{{Heersche2006, moonHo2006, Zyazin2010, SpinHamKim, SpinHamRomeike, SpinHamTimmPRB73_2006, SpinHamTimmMagField2007, SpinHamMisiorny2007, SpinHamLu2009, SpinHamTimm2010, spinFilterBarraza0, spinFilterPark1, Michalak2010, Barraza2010, GgaLdaSanvito}} 

Transport experiments on individual Mn$_{12}$-based SMMs have demonstrated Coulomb blockade, negative differential conductance  and magnetic field dependence of the transport. \cite{Heersche2006,moonHo2006} On the theoretical side,
 two different approaches have been adopted in studying transport through individual SMMs: 
 
 1. Models based on effective spin-Hamiltonians\cite{{Heersche2006, moonHo2006, SpinHamTimmMagField2007, Zyazin2010, SpinHamKim, SpinHamRomeike, SpinHamTimmPRB73_2006, SpinHamMisiorny2007, SpinHamLu2009, SpinHamTimm2010}} have yielded many important insights into the behavior of these systems. They have the advantages of their conceptual simplicity and of the transparency of the results that they yield. However, although in this approach  the spin properties of the SMM are taken into account, the other aspects of the electronic structure of the molecule that play an important role in transport (in particular the spatial details of its molecular orbitals) are not considered. 
 
2. Density functional theory (DFT)-based calculations\cite{PedersonEffectEtralElectron, spinFilterBarraza0, spinFilterPark1, Michalak2010, Barraza2010, GgaLdaSanvito} have overcome this limitation by treating both the spin and spatial electronic structure of a SMM in a unified way. However, DFT transport calculations are time consuming. Thus in the case of large molecules it has been necessary to reduce the number of atoms (by replacing the SMM with a smaller family member \cite{PedersonEffectEtralElectron} or  shortening all or most of the ligands \cite{spinFilterBarraza0,spinFilterPark1,Barraza2010,GgaLdaSanvito}) in order to make the DFT approach to  transport calculations practical.  This simplification, although reasonable, has limitations: Even though the total magnetization of a Mn$_{12}$ SMM covered with CH$_3$ ligands has been predicted not to change if these ligands are removed \cite{Mn12WithoutCH3} or replaced by hydrogen atoms \cite{Mn12WithHydrogenPederson, Mn12WithHydrogen},   the local features of the electronic structure at each Mn site have been predicted to change significantly. \cite{LigandEffectsHan2004}
Furthermore, it has been shown experimentally that ligands can affect the magnetic properties of the SMM.\cite{MnLigandEffectZagaynova} Also, in order to obtain reasonable values of the energy gap between the highest occupied molecular orbital (HOMO) and lowest unoccupied molecular orbital (LUMO) for the SMMs, it has often been necessary to introduce into the DFT-based calculations an adjustable Hubbard U parameter whose value is not known accurately. \cite{spinFilterPark1, GgaLdaSanvito, LDAUBoukhvalov2002, ExpTheoU, LDAUBoukhvalov2004, LDAUPennino, LDAUBarbour}
Thus, while DFT + U calculations have produced more realistic Mn$_{12}$ HOMO-LUMO gaps than pure DFT calculations, the DFT + U calculations carried out by different groups have yielded opposite spin polarizations for the Mn$_{12}$ LUMO orbital.\cite{spinFilterPark1, GgaLdaSanvito}
 
In this paper we describe and analyze a model, that incorporates both the spin and spatial aspects of the electronic structure of SMMs. A preliminary account of the model has been presented elsewhere.\cite{prelim} This model describes the fundamental properties of Mn$_{12}$ SMMs quite well, yielding calculated values of the SMM total spin, the magnetic moments of the inner and outer Mn atoms, the magnetic anisotropy barrier and the HOMO-LUMO gap that are consistent with experiment. However, it is much simpler than DFT and transport computations that are based on it are much less time consuming.  Thus we have been able to investigate two members of the Mn$_{12}$ SMM family including {\em all} of their ligands.  

In our previous paper \cite{prelim} we considered only the neutral Mn$_{12}$-benzoate
single molecule magnet. Here in addition to discussing the electronic and
spintonic structure and transport properties of Mn$_{12}$-benzoate in more detail, we
shall consider also Mn$_{12}$-acetate, both its neutral and negatively charged
species. Mn$_{12}$-acetate and  Mn$_{12}$-benzoate have significantly different
properties. In particular for Mn$_{12}$-acetate the LUMO is located on the Mn core of
the molecule while for Mn$_{12}$-benzoate we predict the LUMO to be located on the
ligands. Therefore comparing and contrasting our results for these two molecules
as we shall do here is of interest. We note that different theories \cite{spinFilterPark1, GgaLdaSanvito} of
Mn$_{12}$-acetate have produced differing results regarding its negatively charged
species and HOMO-LUMO gap. Our theory provides a simple and complete description
of its properties that is consistent with experiment. For neutral and negatively
charged Mn$_{12}$-acetate our tight-binding model yields the total spin of the SMM,
the spins of the individual Mn ions, the magnetic easy axis orientation, the
size of the magnetic anisotropy barrier and the size of the HOMO-LUMO gap
consistent with experiment\cite{Experiment1ElectronEppley}. Therefore, the results that we present here for
Mn$_{12}$-acetate provide a useful benchmark for our tight-binding model.

In the systems considered in the previous  theoretical transport studies of Mn$_{12}$ SMMs the ligands have simply played the role of tunnel barriers and thus the SMMs have always been in the Coulomb blockade regime. Our calculations reveal that this need {\em not} always be the case, with important  consequences for the SMM's transport properties. In particular, for a Mn$_{12}$ SMM covered with 4-(methylthio)benzoate ligands we predict that the LUMO and near LUMO orbitals are localized not on the Mn$_{12}$ core of the molecule (as in the SMMs considered in previous studies) but on some of the ligands. As a consequence we predict that, for some orientations of the SMM relative to gold electrodes and positive gate voltages, the mechanism of conduction through the molecule is off-resonant or resonant tunneling that is {\em not} subject to Coulomb blockade. The HOMO, however is predicted to be localized on the Mn$_{12}$ core of the molecule. As a consequence, for negative gate voltages transport through the SMM is predicted to be in the Coulomb blockade regime. 

In the cases where we predict transport not to be subject to Coulomb blockade the magnetic easy axis of the molecule is oriented approximately parallel to the direction of the current flow through the molecule from one electrode to the other. Thus our theory opens the way to resolving for the first time an important but previously intractable experimental problem, that of determining the direction of the total spin vector of the molecule relative to the electrodes (and to the direction of the current flow): In the Mn$_{12}$ SMM transistors that have been realized experimentally to date, there has been no control over the orientation of the SMM relative to electrodes and how the orientation of the easy axis might be identified experimentally in these systems has remained an open question. We predict that if transport through the molecule is observed to be in the Coulomb blockade regime for negative gate bias but not in the Coulomb blockade regime for positive gate bias, then the easy axis of the molecule is approximately parallel to the direction of the current flow. 

Previous theoretical work\cite{Barraza2010} has suggested that the magnitude of the conductance of a Mn$_{12}$ SMM molecule may depend on the orientation of the easy axis relative to the direction of current flow. We also find the spin-unresolved current through the molecule to depend on the molecule's orientation. However, the conductance also depends on the atomic-scale details of the bonding between the molecule and electrodes. Therefore observation of a {\em qualitative} signature (the molecule switching into and out of the Coulomb blockade regime depending on the sign of the gate voltage) as predicted here would be a much more convincing way to determine the orientation of the easy axis.

It has also been suggested\cite{SpinHamTimmMagField2007} that it may be possible to determine the orientation of the SMM easy axis experimentally by mapping the behavior of transport resonances as functions of the bias voltage and direction of a strong external magnetic field applied to the system. Detecting instead the presence or absence of Coulomb blockade as a function of gate voltage as is proposed here will be more straight forward experimentally. 
 
Our model also predicts strong spin filtering by the SMM in the case of a Mn$_{12}$ molecule in its ground state and strongly coupled to gold electrodes. Whether majority spin (up) or  minority spin (down) electrons are transmitted preferentially depends on both the sign of the gate voltage and on the particular ligands via which the molecule bonds to the electrodes.

In Section \ref{Molecule_Hamiltonian} we introduce our SMM Hamiltonian. Our starting point is the extended H\"{u}ckel model, a tight binding approach that has been used successfully to describe the electronic structures of many non-magnetic molecules.\cite{review2010} We show in this Section how the extended H\"{u}ckel model can be generalized so as to describe the magnetism of SMMs, including the spin polarization and the effects of spin-orbit coupling. In order for the model to describe the SMM magnetic anisotropy barrier correctly, we find it to be necessary to include in it {\em inter}-atomic contributions to the spin-orbit coupling operator as well as the intra-atomic contributions. We derive analytic tight-binding approximations for both contributions in Section \ref{SO} and the Appendix \ref{spin_orbit}. In Section \ref{MoleculeProperties} we present the results of our calculations for the magnetic anisotropy barriers and spin-resolved densities of states for some Mn$_{12}$ SMMs, and examine how these and the nature of the molecular HOMO and LUMO depend on the ligands, the details of the molecular geometry and the molecular charge.
In Section \ref{TransportTheory} we outline the methodology used in our calculations of transport in individual SMMs strongly coupled to the electrodes. We discuss the relationship between the orientation of the SMM easy axis, the SMM molecular orbitals and transport in Section \ref{EasyAxisOrientation}. The lifting of the Coulomb blockade and how this may be used in determining the orientation of the magnetic easy axis are discussed in Section \ref{CB}. We present our predictions regarding spin filtering by Mn$_{12}$ SMMs in Section \ref{SpinFilter}. Our conclusions are presented in Section \ref{Conclusions}.

% ----------------------------------------Section - 2--------------------------------------
\section{The Molecular Model Hamiltonian} 
\label{Molecule_Hamiltonian}

Our SMM Hamiltonian\cite{prelim} is a generalization of the semi-empirical extended H\"{u}ckel model Hamiltonian.\cite{review2010,huckel_off_diagonal,YAEHMOP} It contains the following terms: 
\begin{equation}\label{molecule_ham}
H^{\mbox{\scriptsize{SMM}}}=H^{\mbox{\scriptsize{EH}}}+H^{\mbox{\scriptsize{spin}}}+H^{\mbox{\scriptsize{SO}}}
\end{equation}
Here $H^{\mbox{\scriptsize{EH}}}$ is the extended H\"{u}ckel Hamiltonian,\cite{review2010,huckel_off_diagonal,YAEHMOP} $H^{\mbox{\scriptsize{spin}}}$ gives rise to the spin polarization of the molecule and $H^{\mbox{\scriptsize{SO}}}$ is the spin-orbit coupling term.

\subsection{Extended H\"{u}ckel model} 
\label{EH}

In extended H\"{u}ckel theory, a small set of Slater type atomic valence orbitals $|\Psi_{i\alpha} \rangle$ is chosen as the basis.  Here $|\Psi_{i\alpha} \rangle$ is the $i^\mathrm{th}$ atomic orbital of the $\alpha^\mathrm{th}$ atom. The extended H\"{u}ckel Hamiltonian's diagonal elements $\langle\Psi_{i\alpha} | H^{\mbox{\scriptsize{EH}}}|\Psi_{i\alpha} \rangle =H^{\mbox{\scriptsize{EH}}}_{i\alpha;i\alpha}=\varepsilon_{i \alpha}$ are the experimentally determined negative valence orbital ionization energies $\varepsilon_{i \alpha}$. The non-diagonal matrix elements are given by $H^{\mbox{\scriptsize{EH}}}_{i\alpha;{i}' {\alpha}'}= D_{i\alpha;{i}' {\alpha}'} K \frac{\varepsilon_{i\alpha}+\varepsilon_{{i}' {\alpha}'}}{2}$, where $D_{i\alpha;{i}' {\alpha}'} = \langle \Psi_{i\alpha} |\Psi_{{i}' {\alpha}'} \rangle$ are the orbital overlaps and $K$ is an empirical factor, chosen for
consistency with experimental molecular electronic structure data. In our calculations,\cite{YAEHMOP}    $K=1.75+\Delta_{i\alpha;{i}' {\alpha}'}^{2}-0.75\Delta_{i\alpha;{i}' {\alpha}'}^{4}$ where  $\Delta_{i\alpha;{i}' {\alpha}'}=({\varepsilon_{i\alpha}-\varepsilon_{{i}' {\alpha}'}})/({\varepsilon_{i\alpha}+\varepsilon_{{i}' {\alpha}'}})$ as was proposed in Ref. \onlinecite{huckel_off_diagonal}. 
 
In recent years, extended H\"{u}ckel  theory has been used to understand the transport properties of a variety of non-magnetic molecular systems: Calculations based on extended H\"{u}ckel  theory have yielded tunneling conductances\cite{Datta1997, EmberlyKirczenow01,
Kushmerick02, Cardamone08} and inelastic tunneling intensities\cite{Demir2011} in agreement with experiment for molecules
thiol bonded to gold electrodes. They have also explained 
transport phenomena observed in STM experiments on molecular arrays
on silicon\cite{PivaWolkowKirczenow05, PivaWolkowKirczenow08,PivaWolkowKirczenow09} as well as
electroluminescence data\cite{Buker08}, current-voltage
characteristics\cite{Buker08} and STM images\cite{Buker05} of molecules on
complex substrates. Recently, extended H\"{u}ckel  theory has also been used successfully to model the
electronic structures of adsorbates covalently bonded to graphene and electron transport in graphene
nanoribbons with such adsorbates.\cite{Ihnatsenka11}

\subsection{Spin polarization} 
\label{SP}
 
While extended H\"{u}ckel theory has been used to help construct models describing
spin transport in non-magnetic molecules contacted by magnetic 
electrodes,\cite{EmberlyKirczenow02, DalgleishKirczenow05a, DalgleishKirczenow06a}  
the extended H\"{u}ckel Hamiltonian $H^{\mbox{\scriptsize{EH}}}$ does not 
by itself take account
of spin polarization. It also has only one set of parameters for manganese, 
whereas in the Mn$_{12}$ SMMs the inner and outer Mn ions
have different ($+4, +3$) oxidation states and antiparallel spins.
In our model, $H^{\mbox{\scriptsize{spin}}}$ addresses these issues. 
Its matrix elements $\langle i,s,\alpha |H^{\mbox{\scriptsize{spin}}}| {i}', {s}' ,{\alpha}'\rangle = H^{\mathrm{spin}}_{i s \alpha;{i}' {s}'{\alpha}'}$ between
valence orbitals $i$ and $i'$ of atoms ${\alpha}$ and ${\alpha}'$ with spin $s$ and $s'$ are defined as follows: 
\begin{equation}\label{SpinHamiltonian} \begin{split}
 &H^{\mathrm{spin}}_{i s \alpha;{i}' {s}' {\alpha}'}=D_{i \alpha;{i}'{\alpha}'}\frac{\mathcal{A}_{i \alpha}+\mathcal{A}_{i'{\alpha}'}}{2\hbar} \langle s | \hat{n}\cdot {\bf{S}} | {s}' \rangle \\
 &\mathcal{A}_{i \alpha}={\small \left\{\begin{matrix}
 \mathcal{A}_{\mbox{\scriptsize{inner}}} &\mbox {if $\alpha $ is an inner Mn and $i$ is a {\em d}-valence orbital} \\ 
 \mathcal{A}_{\mbox{\scriptsize{outer}}} &\mbox {if $\alpha $ is an outer Mn and $i$ is a {\em d}-valence orbital} \\ 
 0 &\mbox {otherwise} 
\end{matrix}\right.}
\end{split} \end {equation}
Here $\mathcal{A}_{\mbox {\scriptsize{inner}}}$ and $\mathcal{A}_{\mbox {\scriptsize{outer}}}$ are fitting parameters, $\hat{n}$ is a unit vector aligned with the magnetic moment of the SMM, and  $\bf{S}$ is the one-electron spin operator.
In the Mn$_{12}$ ground state the total spin of each inner Mn ($S_\mathrm{inner}=-\frac{3}{2}$) is antiparallel to the total spin of each outer Mn ($S_{\mbox {\scriptsize{outer}}}=2$).\cite{SMMbookGatteschi} Also the total spin of the Mn$_{12}$ SMM is parallel to the spins of the outer Mn ions and antiparallel to the spins of the inner Mn ions.\cite{SMMbookGatteschi}  Therefore  $\mathcal{A}_{\mbox {\scriptsize{inner}}}$ and $\mathcal{A}_{\mbox {\scriptsize{outer}}}$ have opposite signs ($\mathcal{A}_{\mbox {\scriptsize{inner}}} > 0, \mathcal{A}_{\mbox {\scriptsize{outer}}} < 0 $). Numerical values of these parameters are given in Sec. \ref{MagneticAnisotropyBarrier} and Sec. \ref{LigandEffect}.

\subsection{spin-orbit coupling} 
\label{SO}

Spin-orbit coupling is also not included in extended H\"{u}ckel theory. However, it plays a very important role in SMM physics since it is responsible for the magnetic anisotropy of these systems. We find that in order to describe the SMM magnetic anisotropy correctly within a tight binding model it is necessary to consider the  {\em inter}-atomic contributions to the spin-orbit Hamiltonian operator  $H^{\mbox{\scriptsize{SO}}}$ (perturbation of the atomic orbitals  by the core potentials of {\em other} atoms) as well as {\em intra}-atomic contributions (perturbation of the atomic orbitals by the core potential of the same atom). We derive appropriate analytic expressions for these matrix elements in this paper. Our results generalize extended H\"{u}ckel theory to include the effects of spin-orbit coupling. Spin-orbit coupling in SMMs has previously been treated within the framework of density functional theory.\cite{Mn12WithHydrogenPederson, SpinOrbitDFTKortus, SpinOrbitDFTPark, SpinOrbitDFTSIESTA} The present tight-binding theory has the advantages that it is analytic and relatively straight forward.

The spin-orbit Hamiltonian is given by \cite{Kittel}

{\begin{equation}
\label{SOHamiltonian}
H^{\mbox{\scriptsize{SO}}}=\frac{\hbar}{(2mc)^2}\boldsymbol{\sigma}\cdot  \nabla{V(\bf{r}) \times \mathbf{p}  } 
\end{equation}
where $\mathbf{p}$ is the momentum operator, $V(\bf{r})$ is the electron Coulomb potential energy, $\boldsymbol{\sigma} = (\sigma_x , \sigma_y, \sigma_z) $ and $\sigma_x , \sigma_y $ and $\sigma_z$ are the Pauli spin matrices. 

We approximate $V(\bf{r})$ by a sum of atomic electron potential energies $V(\mathbf{r}) \simeq \sum_{\alpha}{V_ \alpha(\mathbf{r}-\mathbf{r}_\alpha)}$ where $\mathbf{r}_ \alpha $ is the position of $\alpha ^\mathrm{th}$ atomic nucleus. 
Since the spin-orbit coupling arises mainly from the atomic cores where the potential energy $V_ \alpha (\bf{r}-\bf{r}_ \alpha)$ is approximately spherically symmetric we approximate 
$\nabla{V_ \alpha (\bf{r}-\bf{r}_ \alpha) } \simeq  \frac{\bf{r}-\bf{r}_ \alpha}{|\bf{r}-\bf{r}_ \alpha |}\frac{dV_ \alpha(|\bf{r}-\bf{r}_ \alpha |)}{d|\bf{r}-\bf{r}_ \alpha |} $. Then Eq. \ref{SOHamiltonian} becomes
\begin{equation} 
\label{SOHamiltonianstart}
H^{\mbox{\scriptsize{SO}}} \simeq \underset {\alpha}\sum{\frac{1}{2m^2c^2} \frac{1}{|\bf{r}-\bf{r}_ \alpha |}\frac{dV_ \alpha (|\bf{r}-\bf{r}_ \alpha |)}{d |\bf{r}-\bf{r}_ \alpha |}}\mathbf{S}\cdot \mathbf{L}_{\alpha}
\end{equation}
where $\mathbf{S}=\frac{1}{2}\hbar\boldsymbol{\sigma}$ is the electron spin angular momentum operator, ${\bf L}_ \alpha = (\bf{r}-\bf{r}_ \alpha) \times {\bf p}$ is the atomic orbital angular momentum operator with respect to the position of the $\alpha^\mathrm{th}$ nucleus and the sum is over all atoms $\alpha $.
The matrix elements of $H^{\mbox{\scriptsize{SO}}}$ between valence orbitals $i$ and $i'$ of atoms $\beta $ and ${\beta}'$ with spin $s$ and $s'$ are evaluated in Appendix \ref{spin_orbit} and we find

\begin{equation} 
\begin{split}
\label{SOHamiltonianfinal}
\langle {i s \beta}  | H^{\mbox{\scriptsize{SO}}}  | {{i}' s' {\beta}'} \rangle \simeq &
 ~E^{\mbox{\scriptsize{intra}}}_{is{i}'s'; \beta}\delta_{\beta \beta'}+(1-\delta_{\beta \beta'}) \\
&\times \sum_{j}(D_{i \beta; j \beta'}E^{\mbox{\scriptsize{intra}}}_{js i's'; 
\beta'} \\
& +[D_{i' \beta'; j \beta}E^{\mbox{\scriptsize{intra}}}_{js' is; \beta}]^{\ast})
\end{split}
\end{equation}

Here the first term on the right-hand side is the intra-atomic contribution and the remaining terms are the inter-atomic contribution.
An explicit expression for $E^{\mbox{\scriptsize{intra}}}_{is {i}' {s}'; \beta} $, the spin-orbit coupling Hamiltonian matrix element due to {\em intra}-atomic spin-orbit coupling, is Eq. \ref{factorize}.

\section{Properties of Isolated M$\bf{n}_{12}$ Molecules}  \label{MoleculeProperties}

In this Section we present the predictions of the model introduced in Section \ref{Molecule_Hamiltonian}
for isolated SMM molecules that are based on solutions of the tight binding Schr\"{o}dinger equation 
{\begin{equation}  \label{Schrod}
H^ {\mbox{\scriptsize{SMM}}}|\phi_{i}\rangle=E_{i}|\phi_{i}\rangle
\end{equation}}
for the molecular orbitals, $|\phi_{i}\rangle$, and their energies, $E_{i}$. 

It should be noted that in the present work the values of the parameters $H^{\mbox{\scriptsize{EH}}}_{i\alpha,i'\alpha'}$ and $D_{i\alpha,i'\alpha'}$ that enter the extended H\"{u}ckel model, the spin polarization Hamiltonian, Eq. (\ref{SpinHamiltonian}), and the spin-orbit coupling Hamiltonian, Eq. (\ref{SOHamiltonianfinal}), were adopted without modification from  Refs. \onlinecite{huckel_off_diagonal} and \onlinecite{YAEHMOP}.

Experimental estimates\cite{SOMnExperiment} of the spin-orbit coupling constant for manganese, $\epsilon_{{\mbox{\scriptsize{Mn}}}, l_i}$ (see the discussion following Eq. \ref{factorize} in Appendix \ref{spin_orbit}),  have been in the range 0.023--0.051 eV, while theoretical estimates\cite{SOMnTheory} have been in the range 0.038--0.055 eV. In this paper for manganese atoms we use the value $\epsilon_{{\mbox{\scriptsize{Mn}}}, l_i}=$0.036 eV, which is consistent with the experimental and theoretical values. Our calculations show that the spin-orbit coupling constants of the other atoms in molecule do not effect the SMMs' properties significantly. However in the numerical results presented below the spin-orbit coupling due to the O and Au atoms in the extended molecule as well as that due to the Mn atoms is included.
Thus the only free parameters in the present model are $ \mathcal{A}_{\mbox{\scriptsize{inner}}}$ and $ \mathcal{A}_{\mbox{\scriptsize{outer}}}$ of Eq. (\ref{SpinHamiltonian}) that control the spin polarizations of the Mn atoms.

\subsection{Magnetic anisotropy barrier} \label{MagneticAnisotropyBarrier}

The simplest SMM that we shall discuss here is Mn$_{12}$O$_{12}$(O$_{2}$CR)$_{16}$(H$_{2}$O)$_{4}$ with R = CH$_{3}$, henceforth referred to as Mn$_{12}$-Ac. The molecular geometry of Mn$_{12}$-Ac used in the present calculations was obtained from the experimentally measured\cite{Geometry_data_Me} geometry of Mn$_{12}$O$_{12}$(O$_{2}$CR)$_{16}$(H$_{2}$O)$_{4}$ for R = CHCHCH$_{3}$ by substituting CH$_{3}$ for R. 

For the Mn$_{12}$-Ac molecule we chose the values of the magnetic parameters in Eq. \ref{SpinHamiltonian} to be $\mathcal{A}_{\mbox{\scriptsize{inner}}}=$3.0 eV and $\mathcal{A}_{\mbox{\scriptsize{outer}}}=$ -2.6 eV. For these values, we find the calculated ground state total spin of the neutral Mn$_{12}$-Ac to be S$_{\mbox{\scriptsize{total}}}=10$ in agreement with the experimental value.\cite{SpinMn12Me} We find the calculated local magnetic moments associated with inner and outer manganese ions to be -3.18$\mu_{B}$ and 3.86$\mu_{B}$, respectively. According to Hund's rules, these magnetic moments (that correspond to spin $-\frac{3}{2}$ and +2, respectively) are in a good agreement with the Mn$^{4+}$ (Mn$^{3+}$) oxidation states of the inner (outer) manganese ions.\cite{Lis} They are also comparable with the results of other theoretical calculations. \cite{Mn12WithHydrogenPederson, Mn12WithoutCH3}

We estimate the magnetic anisotropy energy of the SMM from the total ground state energy expression
{\begin{equation}  \label{TotalEnergy}
E_{\mbox{\scriptsize{total}}}=\sum_{i}^{}E_{i} -\frac{1}{2} \sum_{i}^{}\left \langle \phi_{i} \left | H^{\mbox{\scriptsize{spin}}} \right | \phi_{i} \right \rangle
\end{equation}}where $E_{i}$ and $|\phi_{i}\rangle$ are molecular orbital (MO) eigenenergies and eigenstates as in Eq. (\ref{Schrod}) and the summations are over all occupied MOs. Since $H^{\mbox{\scriptsize{spin}}}$ represents electron-electron interactions at a mean field level, the second summation on the right-hand side of Eq. (\ref{TotalEnergy}) is required to avoid double counting the corresponding interaction energy. 

\begin{figure}[b!] 
\centering
\includegraphics[width=0.75\linewidth]{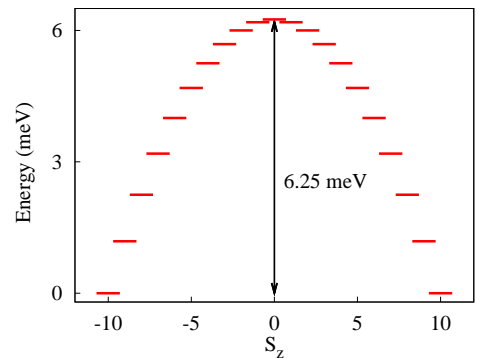} 
 \caption{(Color online) Calculated energy of the neutral Mn$_{12}$-Ac molecule with ground state spin $S_{\mbox{\scriptsize{total}}}=10$ versus the total spin projected on the easy axis. The arrow shows the magnetic anisotropy barrier. } 
\label{MnMeLader} 
\end{figure}

In order to evaluate the magnetic anisotropy barrier (MAB) of the SMM, the total energy of the molecule, Eq.(\ref{TotalEnergy}), has been calculated as a function of the projection of the total spin on the easy axis. As shown in Fig. \ref{MnMeLader}, the calculated MAB of neutral  Mn$_{12}$-Ac is  6.25  meV which is close to its experimental value of  6.1 meV  \cite{expt_MAB}. Our calculations show the contributions of {\em inter}-atomic and {\em intra}-atomic spin-orbit coupling to the MAB to be of the same order of magnitude.}
The results presented above demonstrate that our tight binding model is able to provide a very good description of the experimentally observed magnetic properties of the Mn$_{12}$-Ac SMM.

\subsection{Role of the ligands} \label{LigandEffect}
As has already been mentioned, in DFT-based calculations for SMMs, most or all of the ligands have usually been replaced by hydrogen atoms and thus ligand effects have not been examined fully.\cite{PedersonEffectEtralElectron,GgaLdaSanvito} In order to investigate the ligands' influence on SMM properties, we considered another member of the Mn$_{12}$ family, Mn$_{12}$O$_{12}$(O$_{2}$CC$_{6}$H$_{5}$)$_{16}$(H$_{2}$O$)_{4}$ (referred to as Mn$_{12}$-Ph), that has larger (benzoate) ligands  than Mn$_{12}$-Ac.\cite{prelim}  In the present work the geometry of Mn$_{12}$-Ph has been taken from experimental data.\cite{Geometry_data_Et_and_Ph} 
In our modeling of Mn$_{12}$-Ph we chose the values of the magnetic parameters in Eq. \ref{SpinHamiltonian} to be $\mathcal{A}_{\mbox{\scriptsize{inner}}}=$3.0 eV and $\mathcal{A}_{\mbox{\scriptsize{outer}}}=$-3.5 eV. For these parameter values we find the
calculated ground state's total spin to be S$_{\mbox{\scriptsize{total}}}=10$ and the calculated MAB to be 2.50 meV (see Fig. \ref{MnPhLader} (a)), values that are consistent with experiment. \cite{Geometry_data_Et_and_Ph, MABneutralMn12Ph} 

\begin{figure}[b!] 
\centering
\includegraphics[width=1.0\linewidth]{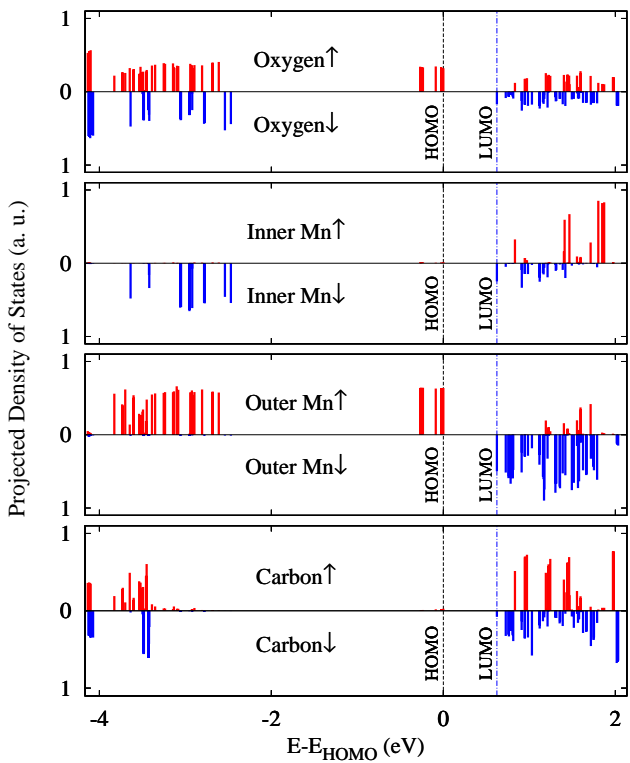} 
\caption{(Color online) Projected density of states for majority spin (up) and  minority spin (down) electrons on the oxygen, inner Mn, outer Mn and carbon atoms
for Mn$_{12}$-Ac.} 
\label{DosMe3} 
\end{figure}

\begin{figure}[b!] 
\centering
\includegraphics[width=1.0\linewidth]{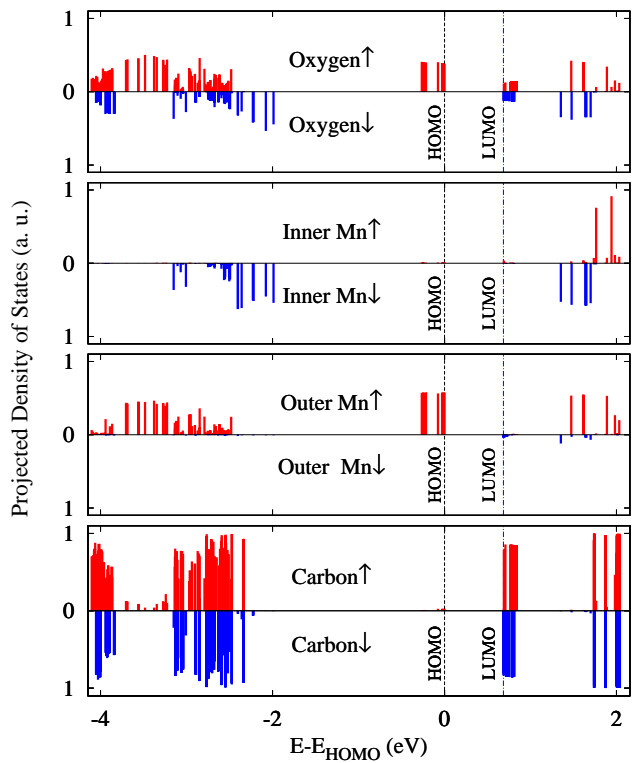} 
\caption{(Color online) Projected density of states for majority spin (up) and  minority spin (down) electrons on the oxygen, inner Mn, outer Mn and carbon atoms
for Mn$_{12}$-Ph.} 
\label{DosMnPh} 
\end{figure}

To compare the effects of the different ligands for Mn$_{12}$, we have plotted the calculated densities of states projected on the oxygen, outer Mn, inner Mn atoms, and carbon atoms in Figs. \ref{DosMe3} and \ref{DosMnPh} for Mn$_{12}$-Ac and Mn$_{12}$-Ph, respectively.
Up to the HOMO energy Fig. \ref{DosMe3} and Fig. \ref{DosMnPh} are similar for both the inner and outer Mn.
In both cases outer Mn are filled mostly with spin up electrons (parallel to the total spin), consistent with having spin S$_{\mbox{\scriptsize{outer-Mn}}}=~$2, the inner Mn are filled with spin down electrons (anti-parallel to the total spin)  in agreement with S$_{\mbox{\scriptsize{inner-Mn}}}=-\frac{3}{2}$, and the carbon atoms are not strongly spin polarized. 
As is seen in in Figs. \ref{DosMe3} and \ref{DosMnPh}, the HOMO level in both cases is predicted to be spin up (majority spin) and located mainly on the outer Mn and on the O atoms, 
consistent with the results of recent 
generalized gradient approximation (GGA) and local density approximation (LDA)+U density functional calculations for Mn$_{12}$-Ac that include the complete set of CH$_{3}$ ligands.\cite{GgaLdaSanvito}

The calculated energy gaps between the HOMO and LUMO (lowest unoccupied molecular orbital) for both the Mn$_{12}$-Ac and Mn$_{12}$-Ph molecules are on the order of 1eV which is common for Mn$_{12}$ family.\cite{MnAcExperimentalMAB}
It is worth mentioning that in DFT calculations, it is typical that LDA and GGA underestimate the HOMO-LUMO gaps in transition metal oxide systems \cite{DFTEnergyGap} and that this has been corrected by using the onsite LDA$+$U correction  where the Hubbard U is treated as an adjustable parameter. For instance the calculated HOMO-LUMO gap for Mn$_{12}$-Ac increased from 0.35 eV in the GGA to 1.6 eV in LDA$+$U in some theoretical studies. \cite{GgaLdaSanvito}

The LUMO can give an indication as to where the added electron may be located in a negatively charged molecule and of the negatively charged molecule's properties. Several experiments have been reported in which reduced anionic products of Mn$_{12}$ have been synthesized in the presence of the PPh$_4$ (PPh$_4$ $\equiv $ P(C$_6$H$_5$)$_4$) molecule which acts as an electron donor and their properties have been studied.\cite{Experiment1ElectronEppley, Experiment1ElectronAubin, Experiment2ElectronSoler} Among the studied molecules are Mn$_{12}$O$_{12}$(O$_{2}$CEt)$_{16}$(H$_{2}$O)$_{4}$ \cite{Experiment1ElectronEppley}, where Et $\equiv $ CH$_2$CH$_3$, (henceforth Mn$_{12}$-Et) and Mn$_{12}$-Ph \cite{Experiment1ElectronAubin}. To our knowledge there has not been any experimental study of a reduced product of Mn$_{12}$-Ac. Since Mn$_{12}$-Et has only one more methylene in its ligands than Mn$_{12}$-Ac, it is reasonable to assume that they have similar properties.

As is seen in Fig. \ref{DosMe3} and \ref{DosMnPh}, our model predicts the LUMOs of both Mn$_{12}$-Ac and Mn$_{12}$-Ph to be spin down electron states, although for Mn$_{12}$-Ph there are also spin up states that are close in energy to the LUMO (see also 
Fig. \ref{MolecularOrbital_lumo} ). Based on this the calculated total spins of the negatively charged Mn$_{12}$-Ac and Mn$_{12}$-Ph molecules are both predicted by the present model to be S$_{\scriptsize{ \mbox{total}}}=9 \frac{1}{2}$, in agreement with experiments. \cite{Experiment1ElectronEppley, Experiment1ElectronAubin}
Although in both cases the LUMOs are spin down states, the predicted locations of the LUMOs for the two molecules are different: In Mn$_{12}$-Ac, an added electron is located primarily on the outer Mn's as is seen in Fig. \ref{DosMe3}, whereas in Mn$_{12}$-Ph, the location of an added electron is primarily on carbon atoms, see Fig. \ref{DosMnPh}. 

It has been found experimentally \cite{Experiment1ElectronEppley} that an added electron on Mn$_{12}$-Et is localized on an outer manganese and the extra electron produces a trapped-valence Mn$^{\scriptsize{\mbox{II}}}$Mn$^{\scriptsize{\mbox{III}}}_{7}$Mn$^{\scriptsize{\mbox{IV}}}_{4}$ system. Our calculation yielded  a similar result for Mn$_{12}$-Ac, where the LUMO is localized primarily on outer manganese atoms. For Mn$_{12}$-Ac, the present model predicts the energy gaps of majority and minority spin to be 0.83 and 3.1 eV, respectively. These values are in good agreement with corresponding GGA estimates \cite{Mn12WithHydrogenPederson, GgaLdaSanvito} of 0.4 and 2.1 eV for majority and minority spin energy gapes, considering that GGA underestimates the HOMO-LUMO gaps in transition metal oxide systems.\cite{DFTEnergyGap} For Mn$_{12}$-Ac the GGA also yields a spin up LUMO primarily on the O and outer Mn while LDA+U yields a spin down LUMO primarily on the inner Mn and O.\cite{GgaLdaSanvito} Interestingly, the result obtained from the GGA\cite{GgaLdaSanvito} for the location of the Mn$_{12}$ LUMO appears to be in better agreement with both the prediction of our model and the above-mentioned experiment\cite{Experiment1ElectronEppley} than that obtained from LDA+U.\cite{GgaLdaSanvito} Therefore, the qualitative features of the HOMO and LUMO of Mn$_{12}$-Ac in the present model agree with experiment on the related Mn$_{12}$-Et molecule, and more consistently than of the DFT--based calculations in the current literature.

For negatively charged Mn$_{12}$-Ph it was found to be too difficult to determine the valence of each manganese ion experimentally.\cite{Experiment1ElectronAubin} Therefore experimental information regarding where the extra electron is localized in this molecule is not available in this time. Our calculation predicts that in Mn$_{12}$-Ph the added electron is localized on the benzoate ligands (including carbon and oxygen atoms) instead of the manganese atoms. To our knowledge
DFT-based calculations of the electronic structure of the Mn$_{12}$-Ph molecule with the complete set of ligands are not available in the literature at this time.

\subsection{Role of geometry}  \label{Geometry Effect}
To investigate the properties of neutral  Mn$_{12}$ SMMs, we have used the available experimental geometries of the neutral Mn$_{12}$ molecules. For negatively charged molecules, as a first approximation, we have assumed in the preceding Section (\ref{LigandEffect}) that the added electron locates where LUMO of the neutral molecule is located, without changing the molecule's geometry. We shall call this the ``unrelaxed charged molecule". By making this idealization we were able to calculate the spin and magnetic anisotropy barrier (MAB) of the negatively charged SMMs.

As has already been mentioned, there have been several experiments studying reduced products of Mn$_{12}.$\cite{Experiment1ElectronEppley, Experiment2ElectronSoler} Therefore as a better approximation, we have also carried out calculations for the Mn$_{12}$-Ph molecule using the experimentally determined geometry of the negatively charged Mn$_{12}$-Ph molecule \cite{Experiment1ElectronAubin}. We will call this geometry the ``realistic charged molecule geometry."

\begin{figure}[t]
\centering
\includegraphics[width=1.0\linewidth]{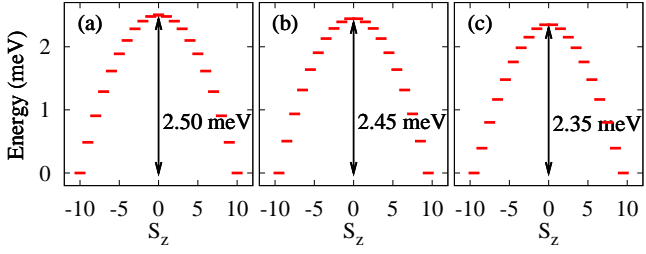}
\caption{(Color online) Calculated energies of Mn$_{12}$-Ph versus total spin projected onto the easy axis for (a) neutral molecule with spin ground state S$_{\mbox{\scriptsize{total}}}=10$ , (b) unrelaxed (explained in text) negatively charged molecule with S$_{\mbox{\scriptsize{total}}}=9\frac{1}{2}$, and (c) realistic (explained in text) negatively charged molecule with S$_{\mbox{\scriptsize{total}}}=9\frac{1}{2}$. Arrows show the magnetic anisotropy barriers.}
\label{MnPhLader}
\end{figure}

Fig. \ref{MnPhLader} (a,b) show the calculated MABs of the neutral and unrelaxed negatively charged Mn$_{12}$-Ph. Even though, according to the experimental data,\cite{Experiment1ElectronAubin} the geometry of the negatively charged Mn$_{12}$-Ph differs from that of the neutral molecule, as is seen in Fig. \ref{MnPhLader} (b) and (c) our calculations did not show a large change in the MAB due to this difference.  This insensitivity of the MAB to the charge state is in agreement with experiment: The experimental values of the lower-temperature  MAB for the neutral \cite{MABneutralMn12Ph} and negatively charged molecule \cite {MABchargedMn12Ph} are 3.3 meV and 2.41 meV, respectively. So, experimentally, the MAB changes by $27\%$  if an extra electron is added.

Our model offers a physical reason for the insensitivity of the MAB to the charge state, as shall be explained next: As is seen in Fig. \ref{DosMnPh} the LUMO of Mn$_{12}$-Ph is  predicted to be located mainly on carbon atoms and not on the manganese. Our calculations yielded a similar result for the realistic charged molecule geometry.  On the other hand the main source of the MAB of Mn$_{12}$ is the Jahn-Teller distortion of the Mn$^{+3}$, see Ref. \onlinecite{moonHo2006}. Therefore even though the geometry has changed due to addition of an extra electron, because the extra electron does not change the oxidation state of the Mn atoms the Jahn-Teller distortion does not change significantly and consequently the MAB does not change by much when an electron is added to the molecule. 

\section{Transport} \label{Transport}
\subsection{Theory of coherent transport in the strong coupling regime} \label{TransportTheory}
 
 In electronic transport experiments the SMM is connected to source and drain leads. Thus, the Hamiltonian for the entire system may be written as:
 \begin{equation}\label{total-ham} 
H=H^{\mbox{\scriptsize{EM}}}+H^{\mbox{\scriptsize{R}}}+H^{\mbox{\scriptsize{L}}}+W^{\mbox{\scriptsize{R}}}+W^{\mbox{\scriptsize{L}}} 
\end{equation}
 Here the system  has been divided into a semi-infinite right lead $H^{\mbox{\scriptsize{R}}}$, a semi-infinite left lead $H^{\mbox{\scriptsize{L}}}$ and an extended molecule $H^{\mbox{\scriptsize{EM}}}$. The extended molecule contains the central SMM as well as a cluster of nearby atoms (assumed here to be gold) belonging to each of the two contacts. We have carried out systematic calculations to determine an adequate gold cluster size for which the transport properties are not affected by the size of the clusters. The results that we present below were obtained for extended molecules with 31 gold atoms in each cluster. There cluster sizes were found to be adequate in the above sense.
 $W^{\mbox{\scriptsize{R(L)}}}$ is the coupling Hamiltonian between the  infinite right(left) lead and extended molecule.

To evaluate $H^{\mbox{\scriptsize{EM}}}$ we have used Eq. (\ref{molecule_ham}) applied to the extended molecule. The right and left leads were modeled as a large number of semi infinite one-dimensional ideal channels that represent macroscopic electron reservoirs, as in previous studies of electron and spin transport through single molecules with gold contacts. \cite{Cardamone10, Cardamone08, Kirczenow07, DalgleishKirczenow06, DalgleishKirczenow05,Demir2011} Nine semi-infinite ideal atomic chains attached to each of the eighteen gold atoms of each cluster that are furthest from the SMM represent the source and drain leads in the transport results that we shall present here.

According to  the Landauer formula, $g=\frac{e^2}{h}T(E_{\mbox{\scriptsize{F}}})$, the conductance is proportional to the total transmission probability $T$ at the Fermi energy $E_{\mbox{\scriptsize{F}}}$. For non-zero temperatures and applied bias voltages $V$, Landauer theory yields\cite{review2010}
\begin{equation}  
	\label{GLCurrent}
 	I(V) = \frac{e}{h} \int_{-\infty}^{\infty} dE~T(E,V) (f(E-\mu_{s})  -  f(E-\mu_{d}) )
\end{equation}
where {\em I} is current,  $f(\varepsilon)$ is the Fermi function, $\mu_{s}$($\mu_{d}$) is the electrochemical potential of the source(drain) electrode.
The total transmission probability can be written as 
\begin{equation}\label{TotalTrasmission} 
{\small T(E,V)= \sum_{ijs{s}'} \frac{v_{j{s}'}}{v_{is}}   \Big\vert t_{j{s}';is}} \Big\vert ^2
\end{equation}
$E$ is the incident electron's energy, $t_{j{s}';is}$ is the transmission amplitude from $i^\mathrm{th}$  electronic channel of left lead with spin $s$  and velocity $v_{is}$ to $j^\mathrm{th}$ electronic channel of the right lead with spin ${s}'$ and velocity $v_{j{s}'}$.

We find the transmission amplitudes {\em{t}} by solving the Lippmann-Schwinger equation 
\begin{equation}\label{LippmannSchwinger} 
\vert \Psi^{\alpha}\rangle =  \vert \Phi_{0}^{\alpha}\rangle +G_0(E) W  \vert \Psi^{\alpha} \rangle 
\end{equation}
where  $\vert \Phi_{0}^{\alpha}\rangle$ is the eigenstate of the $\alpha^\mathrm{th}$ decoupled semi infinite one dimensional lead, $G_0(E)$ is the Green's function of the decoupled system, $W$ is coupling matrix between extended molecule and leads, and $\vert \Psi^{\alpha}\rangle$ is the scattering eigenstate of the coupled system associated with the incident electron state $\vert \Phi_{0}^{\alpha}\rangle$.
Since the basis set used in extended Huckel theory is non orthogonal, we apply the orthogonalization procedure 
described in Ref. \onlinecite{EmberlyKirczenow98} in these calculations.

The Landauer theory described above neglects charging effects that can give rise to Coulomb blockade in a single molecule magnet that is very weakly coupled to {\em all} of the leads that carry electrons to and from the molecule. In some of the cases to be considered below this coupling is found to be {\em very strong} so that Coulomb blockade should not occur and the above Landauer theory is expected to be valid. However, even in the cases where the coupling is weak, calculations based on the Landauer theory provide useful qualitative insights into the molecular states that participate in transport, the degree of broadening of these levels due to their coupling with the leads and hence whether or not Coulomb blockade may be expected to occur.

\subsection{Interplay between the orientation of the magnetic easy axis relative to the leads and electronic structure and transport.}\label{EasyAxisOrientation}

\begin{figure}[b!]
\centering
\includegraphics[width=1.0\linewidth]{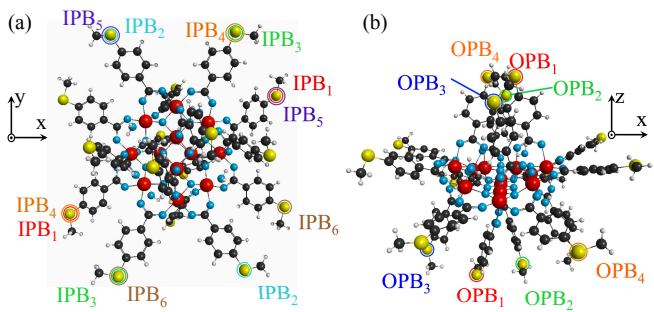}
\caption{(Color online) Two views of the Mn$_{12}$-Ph-Th SMM whose ligands are terminated with methylthio (SCH$_3$) groups. The Mn atoms are located near the $x$-$y$ plane and the magnetic easy axis is aligned with the $z$-axis. We refer to the ligands that are close to the $x$-$y$ plane as ``in-plane ligands" and to the others as ``out-of-plane ligands". The molecule is assumed to be located between two gold leads (not shown) and to bond to each of the gold leads via the sulfur atom of a single ligand, the methyl group having been removed from that sulfur atom. The pairs of open circles with the same color indicate pairs of sulfur atoms that are attached to the gold leads in the different bonding configurations that we consider. (a) The red, blue, green, orange, purple, and brown circles labelled IPB$_i$ indicate the pair of sulfur atoms bonding to the gold in the $ i^\mathrm{th}$ bonding configuration involving in-plane ligands. (b) The red, green, purple, and orange circles labelled OPB$_i$ indicate the pair of sulfur atoms bonding to the gold in the $ i^\mathrm{th}$ bonding configuration involving out-of-plane ligands. Atoms are color labeled: manganese (red), carbon (grey), sulfur (yellow), oxygen (blue), and hydrogen (white).  }
\label{MoleculeViews}
\end{figure}

In the experiments in which transport through a SMM has been measured \cite{Heersche2006, moonHo2006, Zyazin2010} there has been no control over the orientation of the SMM relative to the electrodes. In particular, the orientation of the SMM easy axis has not been controllable. While experimental control over the orientation of the SMM easy axis has not been achieved, it is of interest to consider whether it is possible to experimentally determine the orientation of the easy axis relative to the electrodes that is realized in individual samples by making appropriate measurements.

\begin{figure}[b!] 
\centering
\includegraphics[width=1.0\linewidth]{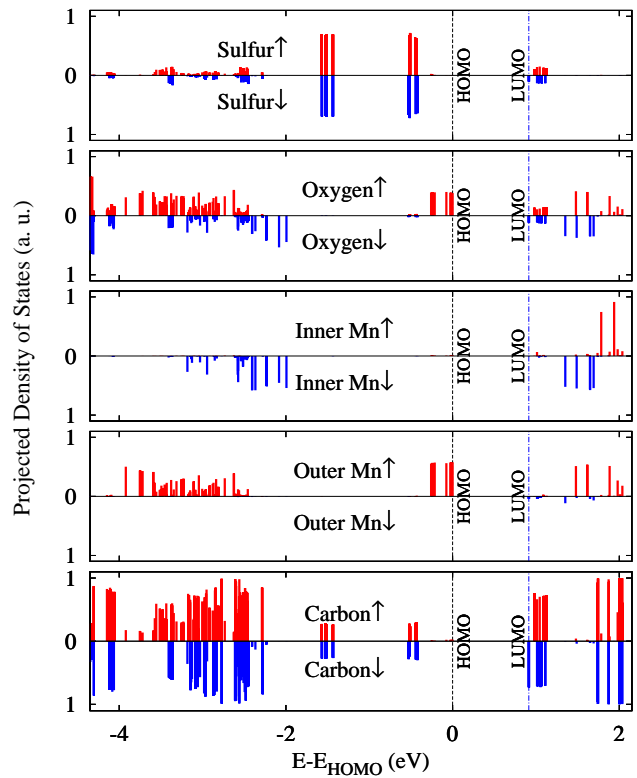} 
\caption{(Color online) Projected density of states for majority spin (up) and  minority spin (down) on the sulfur, oxygen, inner Mn, outer Mn, and carbon atoms for Mn$_{12}$-Ph-Th.} 
\label{DosMnPhTh} 
\end{figure}

\begin{figure}[h!]
\centering
\includegraphics[width=1.0\linewidth]{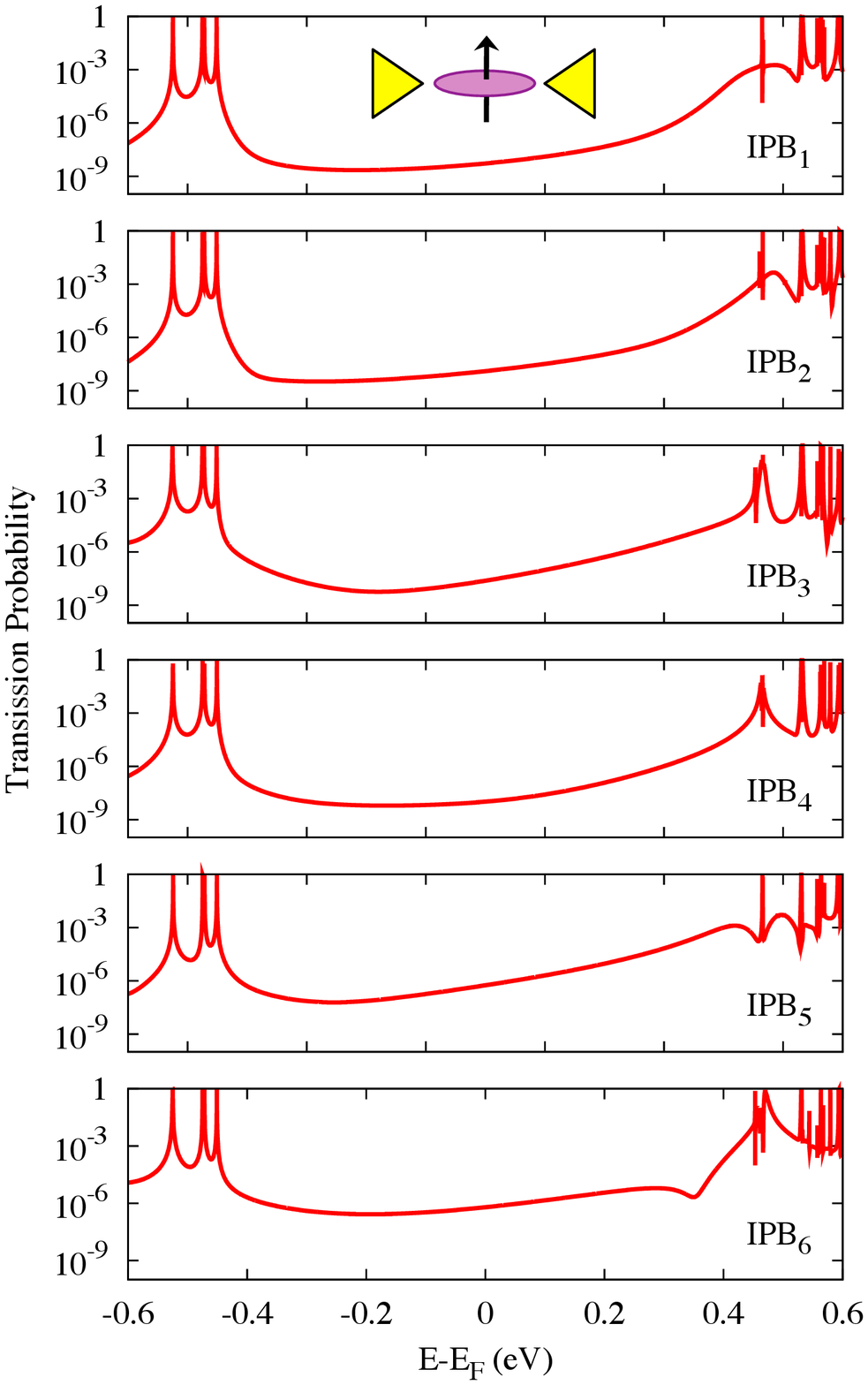}
\caption{(Color online) Total transmission probability at zero bias and gate voltage as a function of energy, $E$, relative to Fermi energy, $E_{F}$, of gold for the different in-plane bonding geometries between the molecule and leads that are identified in Fig. \ref{MoleculeViews} (a). The inset is a schematic of the SMM and electrodes. The magnetic core of the molecule that contains the Mn atoms is pink, the arrow indicates the magnetic easy axis and the electrodes are yellow. }
\label{TransmissionPer}
\end{figure}

\begin{figure}[h!]
\centering
\includegraphics[width=1.0\linewidth]{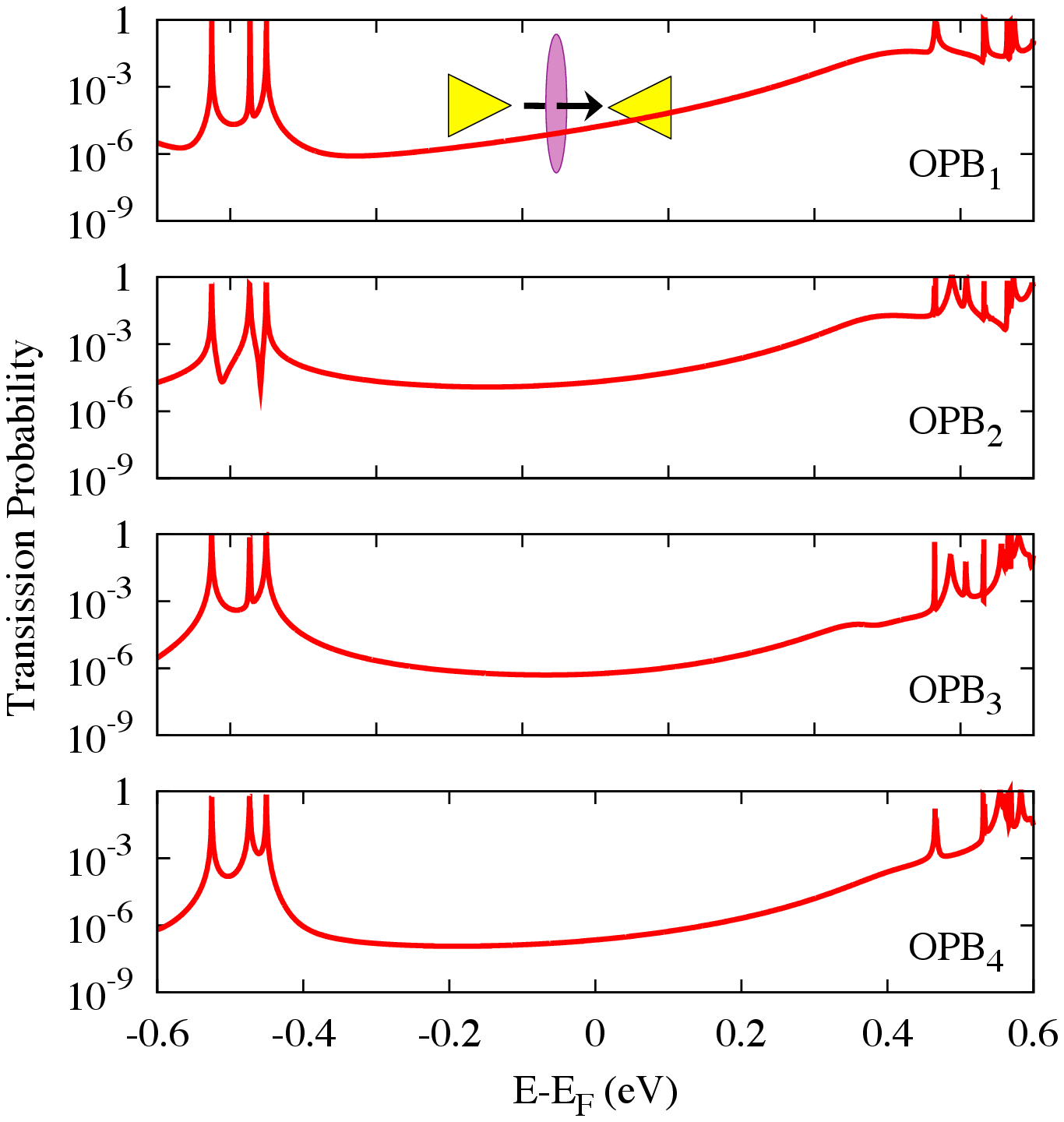}
\caption{(Color online) Total transmission probability at zero bias and gate voltage as a function of energy, $E$, relative to Fermi energy, $E_{F}$, of gold for the different out-of-plane bonding geometries between the molecule and leads that are identified in Fig. \ref{MoleculeViews} (b). The inset is a schematic of the SMM and electrodes. The magnetic core of the molecule that contains the Mn atoms is pink, the arrow indicates the magnetic easy axis and the electrodes are yellow.}
\label{TransmissionPar}
\end{figure}

\begin{figure}[h!]
\centering
\includegraphics[width=0.75\linewidth]{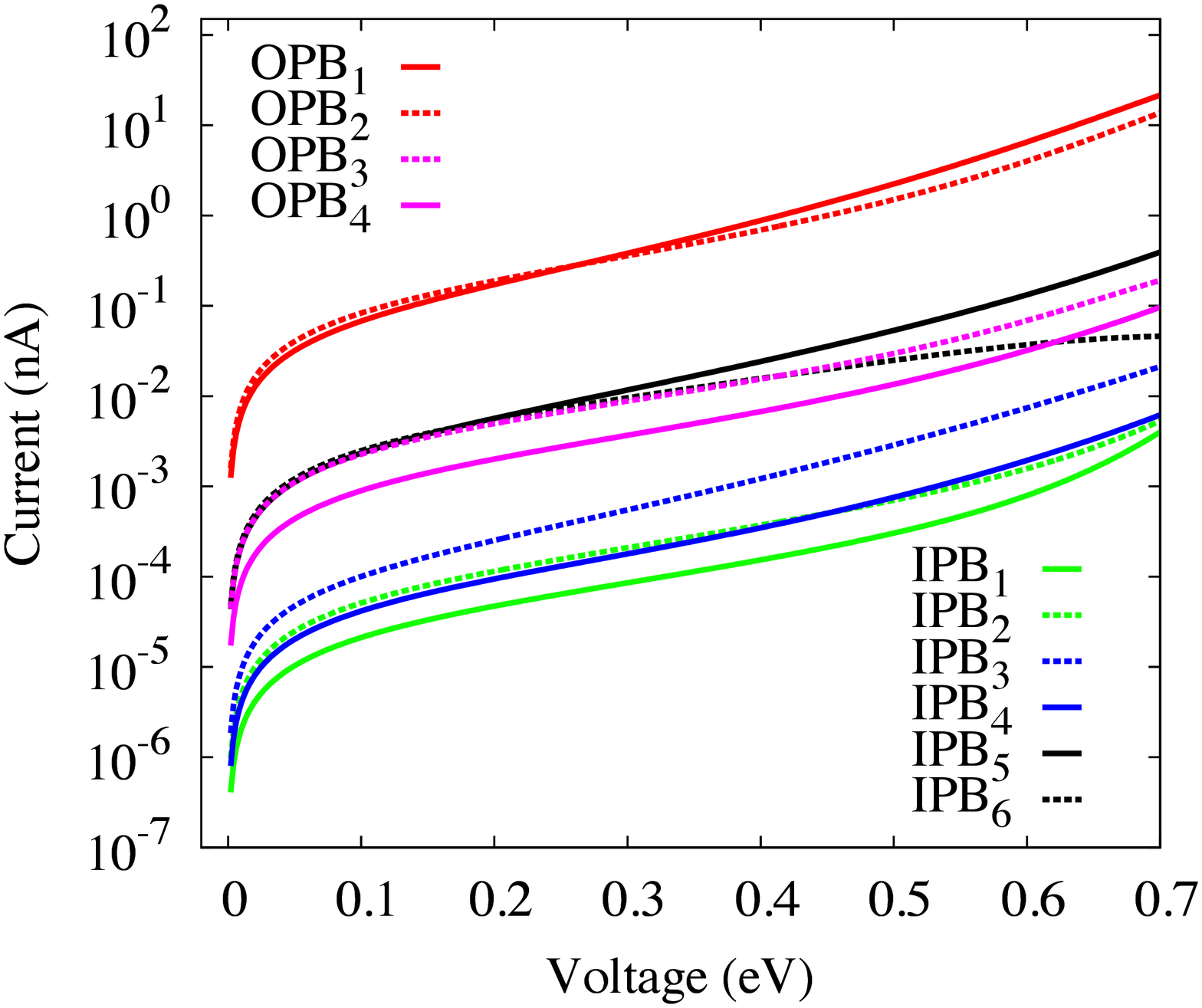}
\caption{(Color Online) Calculated current (from Eq. \ref{GLCurrent}) as a function of bias voltage at zero gate voltage and temperature for different bonding configurations of the SMM relative to electrodes identified in Fig. \ref{MoleculeViews}.}
\label{ComperAllConfiguration}
\end{figure}

We have investigated this by calculating the transport properties of SMMs molecules bridging gold electrodes in a variety of bonding configurations that correspond to different orientations of the easy axis relative to the electrodes.
For our studies of coherent quantum transport through SMMs, we have chosen 4-(methylthio) benzoate Mn$_{12}$ henceforth referred to as Mn$_{12}$-Ph-Th. That is, all of the benzoate ligands are terminated with methylthio (SCH$_3$) groups (see Fig. \ref{MoleculeViews}) except for two ligands that are terminated with sulfur atoms (with their methyl removed) that bond chemically to the gold leads. 

Fig. \ref{DosMnPhTh} shows the projected density of states of Mn$_{12}$-Ph-Th on the sulfur, oxygen, outer Mn, inner Mn, and carbon atoms. The Mn$_{12}$-Ph-Th projected density of states is similar to that of Mn$_{12}$-Ph in Fig. \ref{DosMnPh}. However there are noticeable differences such as the appearance of the sulfur and carbon density of states features below the HOMO in Fig. \ref{DosMnPhTh}, as well as the wave functions of the LUMO and MOs near the LUMO in energy penetrating from the carbon to the sulfur atoms.

If the Mn$_{12}$-Ph-Th molecule shown in Fig. \ref{MoleculeViews} is oriented so that its magnetic easy axis is along the $z$-axis then the Mn atoms are located near the $x$-$y$ plane. With the coordinate axes chosen in this way, we refer to the ligands that are close to the $x$-$y$ plane as ``in-plane ligands" and to the others as ``out-of-plane ligands".
We considered the bonding configurations labelled IPB$_{1}$ - IPB$_{6}$ in Fig. \ref{MoleculeViews} (a) in which the molecule bonds to the leads via different pairs of in-plane ligands and the configurations OPB$_{1}$ - OPB$_{4}$ in Fig. \ref{MoleculeViews} (b) where the molecule bonds to the leads via pairs of  out-of-plane ligands.     

Fig. \ref{TransmissionPer} (Fig. \ref{TransmissionPar}) shows the total electron transmission probability through the molecule as a function of energy, $E$, relative to Fermi energy of electrodes \cite{FermiEenrgy}, $E_{\mbox{\scriptsize{F}}}$, for the different IPB (OPB) configurations.
The details of the calculated transmission spectra for the in-plane-bonding configurations shown in Fig. \ref{TransmissionPer} depend on the ligands involved in the bonding. However, in all cases the transmission probabilities decay very rapidly when the energy falls below that of the LUMO or rises above that of the HOMO. On the other hand, this is {\em not} the case for {\em some} of the out-of-plane-bonding configurations as can be seen in Fig. \ref{TransmissionPar}: In particular, for OPB$_{1}$ and OPB$_{2}$ the transmission peaks near the LUMO show pronounced tails that are at least two orders of magnitude stronger than those for the in-plane bonding configurations in Fig. \ref{TransmissionPer} and extend well into the HOMO-LUMO gap. This difference is manifested in the current carried by the molecule for similar (low and moderate) bias voltages applied across the molecule: As can be seen in Fig. \ref{ComperAllConfiguration}, the calculated currents for the OPB$_{1}$ and OPB$_{2}$ configurations at low and moderate bias are two orders of magnitude larger than for the IPB configurations. On the other hand, the currents for the OPB$_{3}$ and OPB$_{4}$ configurations (for which the tails of the transmission peaks below the LUMO in Fig. \ref{TransmissionPar} are much weaker than for OPB$_{1}$ and OPB$_{2}$) are similar in magnitude to those for some of the IPB configurations. These differences between the transport properties of the SMM in the various bonding geometries may be understood physically as follows:

\begin{figure}[b]
\centering
\includegraphics[width=1\linewidth]{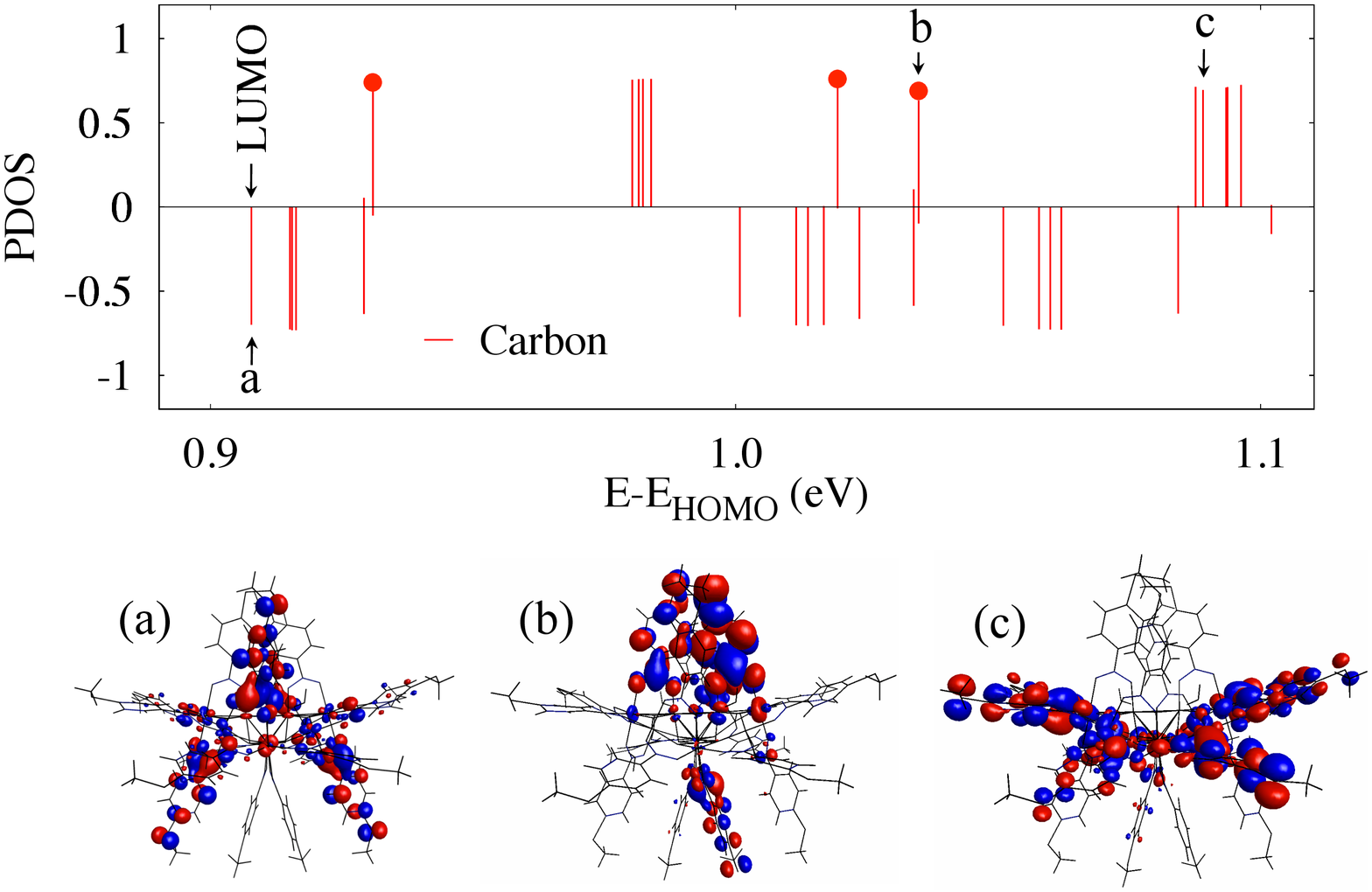}
\caption{(Color Online) Projected density of states(PDOS) on carbon atoms for energies close to the LUMO and related wave function isosurfaces for Mn$_{12}$-Ph-Th. The energies of the states with wave function isosurfaces (a), (b), and (c) are indicated by arrows in the PDOS  plot. The LUMO+5, LUMO+14, and LUMO+17 molecular orbitals are marked in the PDOS plot with filled red circles.} 
\label{MolecularOrbital_lumo}
\end{figure}

The LUMO and the other molecular orbitals that lie closest to it in energy are located mainly on the out-of-plane ligands and have much weaker weight on the in-plane ligands (see orbitals (a) and (b) in Fig. \ref{MolecularOrbital_lumo}). This favors a strong  hybridization of some of the near-LUMO molecular orbitals to gold electrodes that bond to the molecule via out of plane ligands. In cases where this strong hybridization occurs, off-resonant transport {\em at energies below the LUMO} through molecules bound to the electrodes via out-of-plane ligands is expected to be much stronger than that through molecules bound via in-plane ligands. This is consistent with the much larger currents seen in Fig. \ref{ComperAllConfiguration} for OPB$_{1}$ and OPB$_{2}$ configurations than for all of the IPB configurations. Note that the molecular orbitals with larger weight on the in-plane ligands are located further in energy from the LUMO. 
They lie at least 0.18~eV above the LUMO as is shown in Fig. \ref{MolecularOrbital_lumo} (c). Thus off-resonant transport below the LUMO via these orbitals even for gold electrodes bonding to the in-plane ligands is expected to be relatively weak
as is seen in Fig. \ref{ComperAllConfiguration}. 
However, other considerations also play a very important role in determining how well the molecule conducts, as will be discussed next.

 As can be seen in in Fig. \ref{MoleculeViews}, every ligand is attached via oxygen atoms to two Mn atoms. We will refer to these as the ``terminating Mn atoms" of the ligand. We find the magnitude of the current through the molecule at low and moderate bias to be strongly affected by the distances between the terminating Mn atoms belonging to the two ligands that connect the molecule to the electrodes and also by the numbers of oxygen atoms that bond simultaneously to the terminating Mn atoms of those two ligands. These oxygen atoms can be regarded as electronic pathways connecting the two ligands.

For the OPB$_{1}$ geometry two of the terminating Mn atoms of the different ligands are both inner Mn separated by 2.819 {\AA }. In addition to their small separation, these two Mn atoms are connected to each other via two oxygen atoms, i.e., two electronic pathways. 
For OPB$_{2}$ the two ligands that bond to the electrodes share the same (outer) terminating Mn atom. The distance between the other two terminating Mn atoms (one of them is an inner Mn) is also small (2.765  {\AA }), however they are connected to each other through only one oxygen atom (one pathway).
For OPB$_{3}$ all terminating Mn atoms are outer Mn. The smallest distance between two terminating Mn atoms of the different ligands for OPB$_{3}$ is 3.393 {\AA } and these Mn atoms are only connected to each other through one oxygen (one pathway) resulting in the much weaker off resonant transport seen for OPB$_{3}$ than for OPB$_{1}$ and OPB$_{2}$ in Fig. \ref{ComperAllConfiguration}.
For OPB$_{4}$ one of the terminating Mn atoms is an inner Mn. The smallest distance between two terminating Mn atoms of the different ligands for OPB$_{4}$ is 5.44 {\AA } and there is no oxygen atom that bonds simultaneously to these terminating Mn atoms (no direct pathway) resulting in the much weaker off resonant transport seen for OPB$_{4}$ than than for OPB$_{1}$ and OPB$_{2}$ in Fig. \ref{ComperAllConfiguration}. 

\begin{figure}[b]
\centering
\includegraphics[width=1\linewidth]{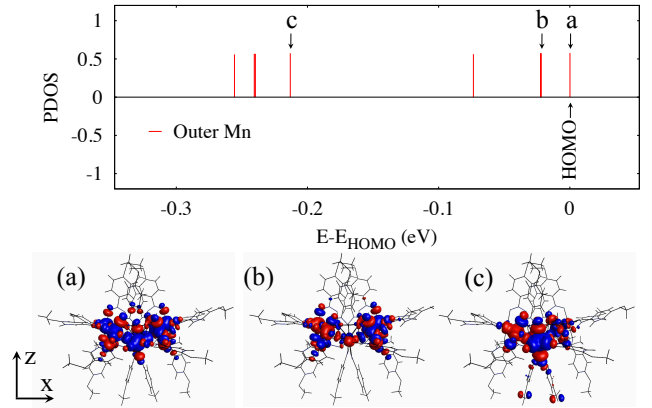}
\caption{(Color Online) Projected density of states(PDOS) on outer Mn atoms in an energy range close to the HOMO and related wave function isosurfaces for Mn$_{12}$-Ph-Th. The energies of the states with wave function isosurfaces (a), (b) and (c) are indicated by arrows in the PDOS plot.}
\label{MolecularOrbital_homo}
\end{figure}

For the in-plane-bonding the terminating Mn atoms are always outer Mn atoms. Among them, IPB$_{5}$ and IPB$_{6}$ have the smallest distances between their terminating Mn atoms, those distances being similar to that for OPB$_{3}$. In addition, for OPB$_{3}$ the terminating Mn are also all outer Mn atoms. These resemblances result in the similar currents for OPB$_{3}$ and IPB$_5$ and IPB$_6$ (see Fig. \ref{ComperAllConfiguration}).

In the transport calculations presented above,  we considered SMMs connected to each gold electrode through one sulfur-gold atomic junction with a sulfur atom at a hollow site (sulfur atom located at the same distance from three gold atoms on a [111] gold surface). In the case of top site bonding (sulfur located at the top of a gold atom) the different bonding geometries have qualitatively similar behavior although the currents are two orders of magnitude smaller than for the hollow site bonding. We have also found currents enhanced by a factor of approximately two if two out-of-plane ligands instead of one bond via sulfur atoms to each electrode\cite{prelim}.

\subsection{Coulomb blockade versus coherent resonant tunneling and the orientation of the magnetic easy axis} \label{CB}

As can be seen in Fig. \ref{MolecularOrbital_lumo} the LUMO of Mn$_{12}$-Ph-Th is located mainly on some of the out of plane ligands of the molecule as are the other molecular orbitals that are closest in energy to the LUMO. By contrast the HOMO seen in Fig. \ref{MolecularOrbital_homo} is located on the central core of the molecule. Thus if the gold electrodes bond chemically to sulfur atoms of the out-of-plane ligands on which the LUMO or/and other MOs close in energy to the LUMO reside then these MOs may hybridize with gold contacts and therefore couple strongly electronically to the gold electrodes. We indeed find this strong hybridization and strong coupling to the electrodes to occur for some of the molecular orbitals that lie close in energy to the LUMO if the molecule bonds to the electrodes in the out-of-plane bonding geometries OPB$_{1}$ and OPB$_{2}$ for which the molecule is the most conductive, as is seen in Fig. \ref{ComperAllConfiguration}. Therefore we predict resonant and off-resonant transport via  these MOs {\em not} to be subject to Coulomb blockade for those bonding geometries. We note that this is a novel prediction of the present model\cite{prelim} as in all of the Mn$_{12}$ SMM systems in previous theoretical transport studies neither the HOMO nor the LUMO had significant presence on any of the ligands. Thus the ligands behaved as tunnel barriers, always resulting in Coulomb blockade of transport.

On the other hand, since in Fig. \ref{MolecularOrbital_homo} the HOMO and orbitals close to it in energy have very little presence on the ligands, we predict that the ligands of Mn$_{12}$-Ph-Th should behave as tunnel barriers in the usual way for transport via the HOMO and nearby orbitals and therefore transport via these orbitals is predicted to be in the Coulomb blockade regime regardless of how the ligands bond to the gold contacts. Furthermore, if the molecule is oriented relative to the electrodes in such a way that only in-plane ligands  bond to the electrodes then transport via the LUMO and nearby orbitals is also predicted to be in the Coulomb blockade regime. Note, however, that for some possible bonding configurations between out-of-plane ligands and the gold electrodes, hybridization between the electrodes and LUMO and near LUMO orbitals need not be strong and/or tunneling through the core of the molecule may be weak. In such cases Coulomb blockade of transport via these orbitals may still occur or transport at energies below the LUMO, while coherent, may still be very weak.

An important consequence of the above predictions is that if in a Mn$_{12}$-Ph-Th transistor transport via the HOMO is observed to be in the Coulomb blockade regime (i.e., Coulomb blockade is observed for negative gate voltages) but transport via the LUMO is observed not to be subject to Coulomb blockade (i.e., Coulomb blockade of the LUMO is not observed for positive gate voltages), then the Mn$_{12}$-Ph-Th molecule is oriented in such a way that at least one out of plane ligand is bonded to an electrode.    

Since the easy axis of the Mn$_{12}$ SMM is roughly aligned with the out-of-plane ligands, this means that it should be possible to identify specific experimental realizations of  Mn$_{12}$-Ph-Th transistors in which the molecular easy axis (and hence the molecular magnetic moment) is approximately aligned with the direction of current flow through the molecule from electrode to electrode simply by observing the presence or absence of Coulomb blockade at the thresholds of conduction for positive and negative gate voltages.

It should be realized, however, that different molecular orbitals that are close in energy to the LUMO may be located primarily on {\em different} out-of-plane ligands. Not {\em all} of the ligands will bond to an electrode in any particular experimental realization of a Mn$_{12}$-Ph-Th transistor. Transport via the molecular orbitals that are close in energy to the LUMO but are located on ligands that have not bonded to an electrode will be subject to Coulomb blockade and its onset will therefore occur at higher positive gate voltages.

\subsection{Spin filtering} \label{SpinFilter}
Previous theoretical work has revealed that Mn$_{12}$ molecules should act as spin filters even when connected to nonferromagnetic gold electrodes.\cite {spinFilterPark1, spinFilterBarraza0, SpinHamTimmPRB73_2006,GgaLdaSanvito} In DFT-based studies at the level of the generalized gradient approximation (GGA),\cite{spinFilterPark1,GgaLdaSanvito,Barraza2010} the Fermi energy of the gold electrodes was found to be between the molecular HOMO and LUMO levels which were both identified as majority spin (up) states. The minority spin (down) MOs were found to be well separated in energy from the majority spin (up) MOs and also from the Fermi energy. On the other hand, DFT-based calculations at the level of LDA + U yielded a majority spin HOMO but either a majority\cite{spinFilterPark1} or minority\cite{GgaLdaSanvito} spin LUMO. However, in each case only majority spin electrons were predicted to be transmitted through the Mn$_{12}$ SMM at low bias voltages. \cite{spinFilterPark1,GgaLdaSanvito,Barraza2010}

As was discussed in Section \ref{MoleculeProperties}, the present model predicts the HOMO to be a majority spin (up) state and the LUMO to be a  minority spin (down) state for Mn$_{12}$-Ac, Mn$_{12}$-Ph, and Mn$_{12}$-Ph-Th, consistent with experiments.\cite{Experiment1ElectronEppley, Experiment1ElectronAubin} This result is also qualitatively consistent with one of the previous LDA + U theoretical studies that predicted a majority spin HOMO and minority spin LUMO for Mn$_{12}$.\cite{GgaLdaSanvito}
However, in the present model the gold Fermi level is located near the center of the molecular HOMO-LUMO gap \cite{FermiEenrgy} for Mn$_{12}$-Ph and Mn$_{12}$-Ph-Th. Thus as will be seen below
spin-filtering with either majority or minority spin electrons being transmitted is predicted to be possible for these molecules, depending on the applied gate voltage and orientation of the molecule.

\begin{figure*}[t]
\centering
\includegraphics[width=1.0\linewidth]{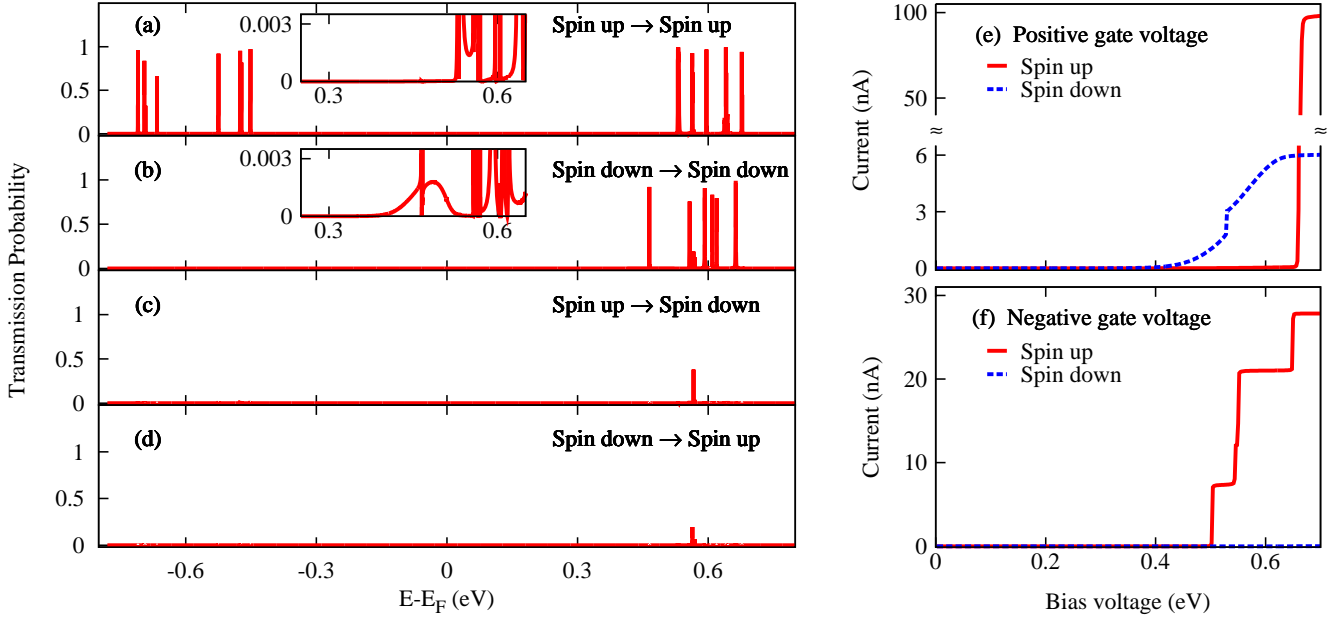}
\caption{(Color online) (a-d) Calculated spin-resolved transmission probabilities at zero bias and gate voltage as a function of energy $E$ relative to Fermi energy $E_{\mbox{\scriptsize F}}$ for a  Mn$_{12}$-Ph-Th molecule bridging a pair of gold electrodes via in-plane ligands in the bonding configuration IPB$_1$ defined in Fig. \ref{MoleculeViews} (a). Transmission of (a) spin up to spin up, (b) spin down to spin down, (c) spin up to spin down,(d) spin down to spin up. Note that the peaks visible in (c) and (d) are a direct consequence of spin-orbit coupling.
The insets show the spin up (down) to spin up (down) transmission probability at zero bias and gate voltage magnified by a factor of 500 relative to the plots in panels (a ) and (b); the tails of the spin down to spin down transmission peaks are responsible for the smooth onset of the spin down current with increasing bias in (e). The calculated spin resolved current is also shown for (e) positive and (f) negative gate voltage as a function of bias voltage at zero temperature. Solid red ( dashed blue) lines represent spin up (down) current. (e) At low bias voltages for positive gate voltages (gate potential at the molecule $=+0.2~$ V) only spin down electrons are transmitted for bias voltages less than 0.65 eV. (f) For the negative gate voltage (gate potential at the molecule $=-0.2~$ V) only spin up electrons are transmitted.}
\label{TransmissionMnPhSIP}
\end{figure*}

In Fig. \ref{TransmissionMnPhSIP}  we present the results of our spin-resolved transport calculations for a particular bonding geometry in which the two in-plane ligands labelled IPB$_1$ in Fig. \ref{MoleculeViews} (a) bond to the gold electrodes. Here the Mn$_{12}$-Ph-Th's easy axis (the $z$-axis in Fig. \ref{MoleculeViews}) is approximately perpendicular to the direction of the current flow between the electrodes and the Mn$_{12}$-Ph-Th molecule is in its ground state (S$=10$ and S$_z=10$). Spin up and spin down are defined relative to the $z$-axis.  

The calculated spin-resolved transmission probabilities at zero source-drain bias are shown in Fig. \ref{TransmissionMnPhSIP} (a-d) as a function of the incident electron energy, $E$, relative to the Fermi energy of the electrodes, $E_{\mbox{\scriptsize F}}$. There is a very close correspondence between peaks of the transmission probabilities in Fig. \ref{TransmissionMnPhSIP} and the spin-resolved density of states in Fig. \ref{DosMnPhTh}. The transmission peaks that are visible at negative energies are due to the HOMO and molecular orbitals (MOs) with lower energies than the HOMO.  Because the MOs close to HOMO are spin up electron states, only spin up electrons are transmitted in this energy range. However, the LUMO and nearby MOs are responsible for peaks at positive energies. As has been discussed above, the LUMO is a spin down state, therefore, the transmission peak with the lowest positive energy corresponds to transmission of only spin down electrons. 

If we regard the gate voltage as an electrostatic potential that shifts the energies of MOs rigidly relative to electrodes, then by applying negative  gate voltages MOs near to HOMO are moved closer to electrode Fermi energy. Then at low bias voltages only spin up electrons are transmitted, i.e. the current through the SMM is purely spin up as is seen in Fig. \ref{TransmissionMnPhSIP}(f). By contrast, if a positive gate voltage is applied, then the transmitted current is spin down at bias voltage below 0.65 eV, Fig. \ref{TransmissionMnPhSIP} (e). Therefore our model predicts that Mn$_{12}$-Ph-Th can be used as a spin filter. 

The smooth rise of the current with increasing bias at low bias that is seen in Fig. \ref{TransmissionMnPhSIP} (e) is due to the low energy tail of the transmission probability close to LUMO that can be seen in the inset of Fig. \ref{TransmissionMnPhSIP} (b), which is due to the weak hybridization of LUMO, LUMO+1, and LUMO+2 with gold contacts. Because the LUMO, LUMO+1, and LUMO+2 have a weak presence on both of the ligands that bond to the gold electrodes  (although the amplitude of the LUMO's wavefunction on these two ligands is {\em much} smaller than that on the out-of-plane ligands) (see Fig. \ref{MolecularOrbital_lumo} (a)) the LUMO, LUMO+1, and LUMO+2 are less weakly coupled electronically to the electrodes than are the other MOs above that LUMO.  However, since the LUMO and MOs that are close to LUMO in energy reside mainly on out-of-plane ligands and the HOMO level resides mainly on the magnetic core of the molecule  these orbitals are all weakly coupled to the leads and therefore should in practice be subject to Coulomb blockade whose main effect will be to suppress the tail of the current at low bias in Fig. \ref{TransmissionMnPhSIP} (e) and to shift the onsets of the current in Fig. \ref{TransmissionMnPhSIP} (e) and (f) to a somewhat higher values of the bias voltage.
As shown in the Fig. \ref{MolecularOrbital_lumo} (c), there are MOs having a strong presence on the in-plane ligands, however they are too far above the LUMO in energy to affect the transport at the low bias voltages shown in Fig.\ref{TransmissionMnPhSIP} (e) and (f) significantly.

As we have mentioned above, the strongly spin-polarized transmission probabilities that give rise to the spin filtering seen in Fig. \ref{TransmissionMnPhSIP} (e-f) occur because of the strong spin polarization of the HOMO and LUMO of the molecule that can be seen in Fig. \ref{DosMnPhTh}. Therefore, qualitatively similar strong spin filtering is also predicted for other in-plane bonding geometries of the molecule and electrodes, provided that the axis of spin quantization is always taken to be the easy axis of the SMM. However, our model predicts the magnitude of the current passing through the molecule to depend very strongly on the bonding geometry.

\begin{figure*}[t]
\centering
\includegraphics[width=1.0\linewidth]{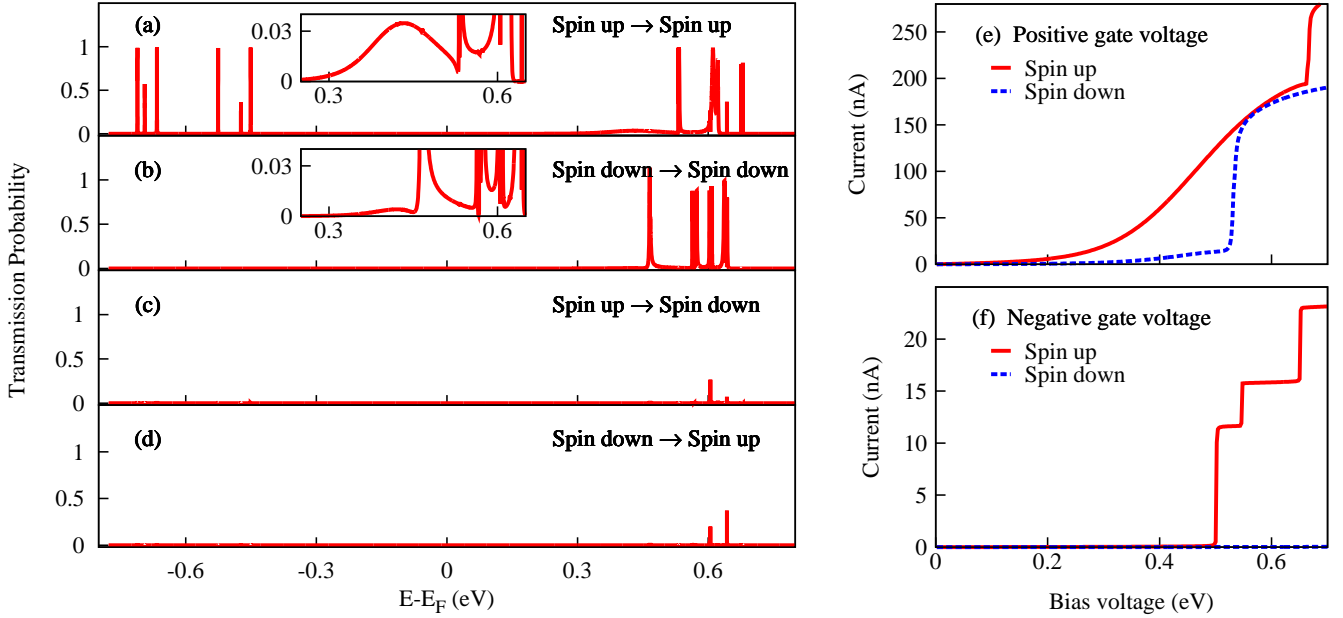}
\caption{(Color online) (a-d) Calculated spin-resolved transmission probabilities at zero bias and gate voltage as a function of energy $E$ relative to Fermi energy $E_{\mbox{\scriptsize F}}$ for a Mn$_{12}$-Ph-Th molecule bridging a pair of gold electrodes via out-of-plane ligands in the bonding configuration OPB$_1$ defined in Fig. \ref{MoleculeViews} (a). Transmission of (a) spin up to spin up, (b) spin down to spin down, (c) spin up to spin down,(d) spin down to spin up.  Note that the peaks visible in (c) and (d) are a direct consequence of spin-orbit coupling. Calculated spin resolved current for (e) positive and (f) negative gate voltage as a function of bias voltage at zero temperature. Solid red (dashed blue) lines represent spin up (down) current. (e) For positive gate voltages (gate potential at the molecule $=+0.2~$ V) at low bias voltage (0 $<$V$_{bias}<$0.53 V) most of the transmitted electrons are spin up. For $V_{bias}>$0.53 V the current is not spin polarized. (f) For negative gate voltages (gate potential at the molecule $=-0.2~$ V) at low bias voltage only spin up electrons are transmitted.}
\label{TransmissionMnPhSOP}
\end{figure*}

In Fig. \ref{TransmissionMnPhSOP}  we present the results of our spin-resolved transport calculations for a particular bonding geometry in which the two out-of-plane ligands labelled OPB$_1$ in Fig. \ref{MoleculeViews} (b) bond to the gold electrodes. Here the Mn$_{12}$-Ph-Th's easy axis (the $z$-axis in Fig. \ref{MoleculeViews}) is approximately parallel to the direction of the current flow between the electrodes and the Mn$_{12}$-Ph-Th molecule is in its ground state (S$=10$ and S$_z=10$). Spin up and spin down are defined relative to the $z$-axis.  

There is again a close correspondence between the peaks of the transmission probabilities in Fig. \ref{TransmissionMnPhSOP} (a-d) and the spin-resolved density of states in Fig. \ref{DosMnPhTh}.

The transmission peaks at low negative energy, due to the HOMO and molecular orbitals with lower energies than the HOMO, for IPB$_1$ and OPB$_1$ are similar (see Fig. \ref{TransmissionMnPhSIP} (a-d) and Fig. \ref{TransmissionMnPhSOP} (a-d)). That is, again only spin up electrons are transmitted in this energy range (see also Fig. \ref{TransmissionMnPhSOP} (f)). 

However, in the OPB$_1$ case in Fig. \ref{TransmissionMnPhSOP} (a) there is a long low energy tail of the spin up to spin up transmission probability that extends far below the LUMO in energy. This tail is due to the strong level broadening that results from the strong hybridization of the LUMO+5, LUMO+14, and LUMO+17 with the gold
contacts.  These MOs are indicated by the solid circles in the projected density of states (PDOS) plot in  Fig. \ref{MolecularOrbital_lumo}.  The hybridization is strong because these orbitals reside mainly on the out-of-plane ligands that are bonded to the electrodes. One of these orbitals is displayed in Fig. \ref{MolecularOrbital_lumo} (b). Even though the LUMO is a spin down state, the off-resonant tunneling through the molecule mediated by this hybridization is responsible for the spin {\em up} current at low bias voltage in Fig. \ref{TransmissionMnPhSOP} (e) since the LUMO+5, LUMO+14, and LUMO+17 are spin up orbitals as is seen in Fig. \ref{MolecularOrbital_lumo}. Note that the LUMO+5, LUMO+14, and LUMO+17 have much more weight on the ligands that connect to the
gold in the OPB$_1$ bonding geometry than does the LUMO. For this reason the hybridization of the spin up LUMO+5, LUMO+14, and LUMO+17 with the gold is much stronger than that of the spin down LUMO and hence the broadening of the LUMO+5, LUMO+14, and LUMO+17 due to their coupling with the gold is also much stronger. Hence the spin up current at low bias voltages in Fig. \ref{TransmissionMnPhSOP} (e) is much stronger than the spin down current.

The very strong hybridization of the LUMO+5, LUMO+14, and LUMO+17 with the gold electrodes in the the OPB$_1$ case and the resulting level broadening  is partly responsible for the low bias tail of the current in Fig. \ref{TransmissionMnPhSOP} (e) being roughly two orders of magnitude stronger than that in Fig. \ref{TransmissionMnPhSIP} (e).
For the same reason, we predict tunneling associated with the LUMO+5, LUMO+14, and LUMO+17 levels for the molecule in the bonding geometry that we are considering in Fig. \ref{TransmissionMnPhSOP} to be immune to suppression by charging effects, i.e., there should be no Coulomb blockade of transport associated with the LUMO+5, LUMO+14, and LUMO+17. By contrast, since the HOMO and nearby levels reside mainly on the magnetic core of the molecule they are much more weakly coupled to the leads and much less broadened. This is evident from the  lack of visible tails of the resonant peaks at negative energies in Fig. \ref{TransmissionMnPhSOP} (a) and the correspondingly very abrupt character of the current steps in Fig. \ref{TransmissionMnPhSOP} (f).  Therefore transport via the HOMO and nearby levels should be subject to Coulomb blockade whose main effect will be to shift the onset of the current in Fig. \ref{TransmissionMnPhSOP} (f) to a somewhat higher value of the bias voltage. 

If we regard the gate voltage as an electrostatic potential that shifts the energies of MOs rigidly relative to electrodes, then applying negative  gate voltages shifts MOs that are near to HOMO closer to the electrode Fermi energy. Then at low bias voltages purely spin up electrons are transmitted, similarly to the case of IPB$_1$ bonding to the electrodes. I.e. the current through the SMM is purely spin up as is seen in Fig. \ref{TransmissionMnPhSOP} (f). But, in contrast to IPB$_1$, if a positive gate voltage is applied, then the transmitted current at low bias voltage (0 $<$ V$_{bias} <$ 0.53 V) is predominantly  spin up, Fig. \ref{TransmissionMnPhSOP} (e). However, at bias voltages above 0.53 V, when the LUMO and nearby MOs which have spin down become available for transport, the transmitted current ceases to be spin polarized.

Therefore our model predicts that Mn$_{12}$-Ph-Th can be used as a spin filter. However, we predict that it should exhibit differing behaviour depending on the orientation of molecule relative to the electrodes. At negative gate voltages for both in-plane bonding and out-of-plane bonding configurations we predict that the molecule should transmit spin up electrons. However, for positive gave voltages at low bias voltage, in the case of in-plane bonding geometries we predict spin down electrons to be transmitted, whereas for out-of-plane bonding we predict spin up electrons to be transmitted. For positive gave voltages at higher bias voltages the transmitted current ceases to be spin polarized.

Note that although the transport calculations presented in this work do not include the effects of Coulomb blockade explicitly whereas transport via the molecular HOMO and near HOMO levels is expected to be subject to Coulomb blockade, our prediction of effective spin filtering occurring in transport via the molecular HOMO and near HOMO levels is expected to be valid in the presence of Coulomb blockade {\em at the Coulomb blockade threshold}. This is because of at the Coulomb blockade threshold the only available state for electron to pass through the molecule is the HOMO or a near HOMO orbital   all of which are predicted to have spin aligned with easy axis \cite{Michalak2010}(see Fig. \ref{MnPhLader} (a,c)). Therefore spin flip processes (that would degrade the spin filtering) are not permitted energetically on the molecule during transport via the HOMO and near HOMO levels even for incoherent transport in the Coulomb blockade regime.

\section{Conclusions} \label{Conclusions}

In this paper we have introduced a tight-binding model of single molecule magnets that incorporates both the spin and spatial aspects of the molecular electronic structure. It is able to provide a realistic description of the known properties of Mn$_{12}$ single molecule magnets for both the neutral and negatively charged molecules. The properties calculated within the present model that are consistent with the available experimental data include the total spin of the molecule, the spins of the individual Mn ions, the magnitude of the magnetic anisotropy barrier and its orientation and the size of the HOMO-LUMO gap. The spins and spatial locations in the molecule of the HOMOs predicted by the model are consistent with the results of density functional theory based calculations available in the literature.\cite {PedersonEffectEtralElectron, spinFilterBarraza0, spinFilterPark1, GgaLdaSanvito, Mn12WithHydrogenPederson} Different density functional theory based calculations have yielded differing results for the spins and spatial locations of the LUMOs.\cite {spinFilterPark1, GgaLdaSanvito} The predictions of the present model for  the spins and spatial locations of the LUMOs are consistent with experimental data for the corresponding negatively charged molecules. 

The description of the properties of the Mn$_{12}$ family of single molecule magnets provided by the model is comparable to and in some respects more realistic than that provided by density functional theory that has been improved by the inclusion of a  phenomenological Hubbard U parameter. However, calculations based on the present model are much simpler and require far less computational resources and compute time than density functional theory-based calculations. We have therefore been able to investigate the electronic and spintronic structure and transport properties of more complete models of Mn$_{12}$ single molecule magnets than have been studied previously. In particular, we have presented the results of the first transport calculations for Mn$_{12}$ single molecule magnets with complete sets of ligands none of which have been truncated or replaced with hydrogen atoms.

In the models of transport in Mn$_{12}$ single molecule magnets that have been studied previously theoretically, the ligands have always behaved as simple tunnel barriers between the magnetic core of the molecule and the electrodes and thus transport has been in the Coulomb blockade regime. Here we have shown that for Mn$_{12}$ single molecule magnets covered with thiolated or methylthio terminated aromatic benzoate ligands, the situation can be very different: We predict that for these molecules the molecular LUMO and molecular orbitals close in energy to the LUMO reside on some of the ligands. This has important novel implications for the transport properties of these single-molecule magnets: In the case where a molecular orbital resides on one or more ligands that bond to an electrode, we predict low temperature conduction via that molecular orbital to be in the coherent resonant or off-resonant quantum regime and {\em not} subject to Coulomb blockade that has until now been assumed to always govern electrical conduction through Mn$_{12}$ single-molecule magnets.    Because we find the LUMO and nearby orbitals to be localized on ligands that are aligned approximately parallel to the magnetic easy axis of the Mn$_{12}$ single molecule magnet, we propose that observation of the absence of Coulomb blockade at low source-drain bias and positive gate voltages can be used as an experimental signature identifying specific realizations of single molecule magnet transistors in which the magnetic easy axis and magnetic dipole moment are aligned approximately parallel to the direction of current flow through the molecule. We also predict that Mn$_{12}$ single molecule magnets covered with thiolated aromatic benzoate ligands should behave as effective spin filters, and that this should be the case even for transport at low bias via the molecular orbitals that reside primarily on the ligands and not on the magnetic core of the molecule. While we have predicted that the LUMO should lie on the ligands of Mn$_{12}$-benzoate and its derivatives, it is reasonable to expect the LUMO and/or the HOMO to lie on the ligands of some other Mn$_{12}$ single molecule magnets as well. Examples of such systems may be Mn$_{12}$ with ligands derived from polyacetylene and polythiophene that have small HOMO-LUMO gaps, 1.4 eV\cite{polyacetylene} and 0.85 eV\cite{polythiophene}, respectively. 

If these predictions are confirmed experimentally, the implications for the field of transport in single molecule magnets will be significant: Transport through Mn$_{12}$ single molecule magnets that is {\em not} in the Coulomb blockade regime will have become experimentally accessible. Also, single molecule magnet transistors with a known orientation of the magnetic easy axis (and hence a known orientation of the spins carried by the spin-filtered current) relative to the direction of current flow through the single molecule magnet will have been realized.

\begin{acknowledgments}
This research was supported by CIFAR, NSERC, Compute Canada and Westgrid. 
We thank  B. Gates, B. L. Johnson and A. Saffarzadeh for helpful comments and discussions.
\end{acknowledgments}
 
\appendix
\section{} \label{spin_orbit}
In this Appendix we evaluate the matrix elements of the spin-orbit coupling Hamiltonian $H^{\mbox{\scriptsize{SO}}}$ between valence orbitals $i$ and $i'$ of atoms $\beta$ and ${\beta}'$ with spin $s$ and $s'$. Eq. \ref{SOHamiltonianstart} of Section \ref{SO} in matrix form is

{\scriptsize 
\begin{equation} \begin{split} \label{GSP1}
E^{\mbox{\scriptsize{so}}}_{i s \beta;{i}' {s}' {\beta}'} &= \langle \Psi_{i s \beta} | H^{\mbox{\scriptsize{SO}}} | \Psi_{{i}' {s}' {\beta}'} \rangle \\
&\simeq\underset {\alpha}\sum\langle \Psi_{i s \beta} | 
{\frac{1}{2m^2c^2} \frac{1}{|\bf{r}-\bf{r}_\alpha|}\frac{dV_\alpha (|\bf{r}-\bf{r}_\alpha|)}{d |\bf{r}-\bf{r}_\alpha|}}
\mathbf{S}\cdot \mathbf{L}_{\alpha}| \Psi_{{i}' {s}' {\beta}'} \rangle
 \end{split} \end{equation} }where $\Psi_{i s \beta}$ is the $i^\mathrm{th}$ atomic orbital of the $\beta^\mathrm{th}$ atom with spin $s$. 
The largest terms in the summation in Eq. \ref{GSP1} are those in which (atom) $\alpha$ is the same as $\beta$ 
or ${\beta}'$ or both. In what follows, we shall consider only these terms.  

Then for $\beta={\beta}'$, the spin-orbit coupling Hamiltonian matrix element is due to {\em intra}-atomic spin-orbit coupling. We will refer to it as $E^{\mbox{\scriptsize{intra}}}_{is {i}' {s}';\beta} \equiv E^{\mbox{\scriptsize{so}}}_{i s \beta;{i}' {s}' \beta}$~. The atomic orbital wave function $\Psi_{i s \beta}$   can be expressed as the product of a radial wave function $R_{\beta, l_i}$ and directed atomic orbital $|\beta, l_i, d_i, s\rangle$. Here $l_i$ is the angular momentum quantum number, $d_i$ may be $s, p_x, p_y, p_z, d_{xy},d_{xz},...$ depending on the value of $l_i$, and $s$ is the spin quantum number. Then
\begin{widetext}
{\begin{equation}    \label{factorize}
E^{\mbox{\scriptsize{intra}}}_{i s {i}' {s}';\beta} =  \langle \beta, l_i, d_i, s | \mathbf S \cdot \mathbf L_{\beta} | \beta, l_{i'}, d_{i'}, {s}' \rangle  
\langle R_{\beta, l_i}  | \frac{1}{2m^2c^2} \frac{1}{|\mathbf{r}-\mathbf{r}_{\beta}|} \frac{dV(|\mathbf{r}-\mathbf{r}_{\beta}|)}{d(|\mathbf{r}-\mathbf{r}_{\beta}|)}  | R_{\beta, l_{i'}}  \rangle
 \end{equation}}

The ${\bf S} \cdot \mathbf L_{\beta}$ operator does not change the angular momentum quantum number $l_{i'}$, therefore $E^{\mbox{\scriptsize{intra}}}_{i s {i}' {s}';\beta}$ is non-zero only when $l_{i}=l_{i'}$. Explicit expressions for the resulting matrix elements of ${\bf S} \cdot \mathbf L_{\beta}$ in Eq. \ref{factorize} have been given in Ref. \onlinecite{SpinOrbiGraphene}. 
The radial integrals, $\langle R_{\beta, l_i}  | \frac{1}{2m^2c^2} \frac{1}{|\mathbf{r}-\mathbf{r}_{\beta}|} \frac{dV(|\mathbf{r}-\mathbf{r}_{\beta}|)}{d(|\mathbf{r}-\mathbf{r}_{\beta}|)}  | R_{\beta, l_{i}}  \rangle = \epsilon_{\beta, l_i}$,  are the spin-orbit coupling constants which are available in the literature for various atoms and ions. The numerical values of these constants are discussed in Sec. \ref{MoleculeProperties}.  

For the {\em inter}-atomic contributions to the spin-orbit coupling, in Eq. \ref{GSP1} $\beta \ne \beta'$  and the summation over $\alpha$ reduces to two dominant terms, those for which $\alpha=\beta'$ or $\alpha=\beta$. The term $\alpha=\beta'$ is

{\small
\begin{eqnarray} 
&&\langle \Psi_{i s \beta} | 
{\frac{1}{2m^2c^2} \frac{1}{|\bf{r}-\bf{r}_{\beta'}|}\frac{dV_{\beta'} (|\bf{r}-\bf{r}_{\beta'}|)}{d |\bf{r}-\bf{r}_{\beta'}|}}
\mathbf{S}\cdot \mathbf{L}_{{\beta'}}| \Psi_{{i}'  {s}'{\beta'}} \rangle \label{iEuqalToq1} \\
&&=\underset {j}\sum\langle \Psi_{i s \beta} |  \Psi_{j s_j \beta'} \rangle \langle \Psi_{j s_j \beta'}|
{\frac{1}{2m^2c^2} \frac{1}{|\bf{r}-\bf{r}_{\beta'}|}\frac{dV_{\beta'} (|\bf{r}-\bf{r}_{\beta'}|)}{d |\bf{r}-\bf{r}_{\beta'}|}}
\mathbf{S}\cdot \mathbf{L}_{{\beta'}}| \Psi_{{i}' {s}' {\beta'}} \rangle \label{iEuqalToq2}  
\end{eqnarray}}

\end{widetext}
The summation in expression \eqref{iEuqalToq2} is over all of the atomic orbitals of the $\beta'^\mathrm{th}$ atom that form a complete orthogonal set. However, since our goal is to construct a generalization of extended H\"{u}ckel theory that includes spin-orbit coupling we retain in the sum only the atomic {\em valence} orbitals. Then the expression  \eqref{iEuqalToq2} reduces to $\sum_{j}D_{i \beta; j\beta'}E^{\mbox{\scriptsize{intra}}}_{j s i'  {s}'; \beta'}$ where $D_{i \beta; j\beta'}=\langle\Psi_{i \beta} |\Psi_{j \beta'} \rangle \delta_{s, s_j}$.

The remaining term $\alpha=\beta$ of Eq. (\ref{GSP1}) is 
{\small
\[ \begin{split}
&\langle \Psi_{i s \beta} | 
{\frac{1}{2m^2c^2} \frac{1}{|\bf{r}-\bf{r}_\beta|}\frac{dV_\beta (|\bf{r}-\bf{r}_\beta|)}{d |\bf{r}-\bf{r}_\beta|}}
\mathbf{S}\cdot \mathbf{L}_{\beta}| \Psi_{{i}' {s}' \beta'} \rangle  \\
&=\langle \Psi_{{i}' {s}' \beta'} | 
{\frac{1}{2m^2c^2} \frac{1}{|\bf{r}-\bf{r}_\beta|}\frac{dV_\beta (|\bf{r}-\bf{r}_\beta|)}{d |\bf{r}-\bf{r}_\beta|}}
\mathbf{S}\cdot \mathbf{L}_{\beta}| \Psi_{i s \beta} \rangle ^\ast\\
\end{split} \]}since ${\bf S} \cdot \mathbf L_{\beta}$ is Hermitian. 
The last expression is evaluated in a similar way to Eq. \eqref{iEuqalToq1}. 
Finally, collecting the above results, Eq. (\ref{GSP1}) for the matrix elements of the spin-orbit coupling Hamiltonian reduces to

\begin{equation} 
\begin{split}
\label{SOHamiltonianfinal}
\langle {i s \beta}  | H^{\mbox{\scriptsize{SO}}}  | {{i}' s' {\beta}'} \rangle \simeq &
 ~E^{\mbox{\scriptsize{intra}}}_{is{i}'s'; \beta}\delta_{\beta \beta'}+(1-\delta_{\beta \beta'}) \\
&\times \sum_{j}(D_{i \beta; j \beta'}E^{\mbox{\scriptsize{intra}}}_{js i's'; 
\beta'} \\
& +[D_{i' \beta'; j \beta}E^{\mbox{\scriptsize{intra}}}_{js' is; \beta}]^{\ast})
\end{split}
\end{equation}Here the first term on the right-hand side is the intra-atomic contribution and the remaining terms are the inter-atomic contribution.

%% ----------------------------------------------------------------------------------------------------
{

\end{document}
\begin{thebibliography}{333}
{\footnotesize 

\bibitem{SMM2008} L. Bogani and W. Wernsdorfer, Nature Mater. {\bf 7}, 179 (2008)
\bibitem{SMMbookGatteschi}D. Gatteschi, R. Sessoli, and J. Villain, \textit{Molecular Nanomagnets} (Oxford University Press, New York, 2006).
\bibitem{SMMChristou} G. Christou, Polyhedron {\bf 24}, 2065 (2005)
\bibitem{SMMGatteschi} D. Gatteschi, R. Sessoli, Angew. Chem. Int. Ed. {\bf 42}, 268 (2003).




\bibitem{Lis} T. Lis, Acta Cryst. B {\bf 36}, 2042 (1980)

\bibitem{S10} A. Caneschi, D. Gatteschi, R. Sessoli, A. L. Barra, L. C. Brunel, M. Guillot, J. Am. Chem. Soc. {\bf 113}, 5873 (1991)


\bibitem{expt_MAB} A. L. Barra, D. Gatteschi, R. Sessoli, Phys. Rev. B {\bf 56}, 8192 (1997).

\bibitem{MnLigandEffectZagaynova} V. S. Zagaynova, T. L. Makarova, N. G. Spitsina, D. W. Boukhvalov, Journal of Superconductivity and Novel Magnetism, {\bf 24}, 855 (2011). 


\bibitem{expt_RelaxationTime}  R. Sessoli, D. Gatteschi, A. Caneschi, and M. A. Novak, Nature{\bf  365} , 141 (1993).

\bibitem{review2010}  For a recent review see G. Kirczenow, \textit{Molecular
nanowires and their properties as electrical conductors}, The Oxford Handbook
of Nanoscience and Technology,Volume I: Basic Aspects, Chapter 4, edited by A.
V. Narlikar and Y. Y. Fu, Oxford University Press, U.K. (2010).

\bibitem{Heersche2006} H. B. Heersche, Z. de Groot, J. A. Folk, H. S. J. van der Zant, C. Romeike, M. R. Wegewijs, L. Zobbi, D. Barreca, E. Tondello, and A. Cornia, Phys. Rev. Lett. {\bf 96} 206801 (2006). 

\bibitem{moonHo2006} M.-H. Jo, J. E. Grose, K. Baheti, M. M. Deshmukh, J. J. Sokol, E. M. Rumberger, D. N. Hendrickson, J. R. Long, H. Park, and D. C. Ralph, Nano Lett. {\bf 6}, 2014 (2006).

\bibitem{Zyazin2010} A. S. Zyazin, J. W. G. van den Berg, E. A. Osorio, H. S. J. van der Zant, N. P. Konstantinidis, M. Leijnse, M. R. Wegewijs, F. May, W. Hofstetter, C. Danieli, and A. Cornia, Nano Lett. {\bf 10}, 3307 (2010).

\bibitem{SpinHamKim} G.-H. Kim and T.-S. Kim, Phys. Rev. Lett. {\bf 92}, 137203 (2004).
\bibitem{SpinHamRomeike}C. Romeike, M. R. Wegewijs, H. Schoeller,  Phys. Rev. Lett. {\bf 96}, 196805 (2006).  
\bibitem{SpinHamTimmPRB73_2006} C. Timm, and F.  Elste, Phys. Rev. B {\bf 73}, 235304 (2006).

\bibitem{SpinHamTimmMagField2007} C. Timm, Phys. Rev. B {\bf76}, 014421 (2007).

\bibitem{SpinHamMisiorny2007} M. Misiorny, I. Weymann, J. Barnas,  Phys. Rev. B 79, 224420 (2009).

\bibitem{SpinHamLu2009}H.-Z. Lu, B. Zhou, and S.-Q. Shen, Phys. Rev. B {\bf 79}, 174419 (2009).
\bibitem{SpinHamTimm2010} F. Elste, C. Timm,  Phys. Rev. B {\bf 81}, 024421 (2010).


\bibitem{spinFilterBarraza0}S. Barraza-Lopez, K. Park, V. Garc\'{i}a-Su\'{a}rez and J. Ferrer, J. Appl. Phys. {\bf 105}, 07E309 (2009).

\bibitem{spinFilterPark1} S. Barraza-Lopez, K. Park, V. Garc\'{i}a-Su\'{a}rez, and J. Ferrer, Phys. Rev. Lett. {\bf 102}, 246801 (2009). 

\bibitem{GgaLdaSanvito}C. D. Pemmaraju, I. Rungger, S. Sanvito, Phys. Rev. B {\bf 80}, 104422 (2009).



\bibitem{Barraza2010} K. Park, S. Barraza-Lopez, V. M. Garc\'{i}a-Su\'{a}rez, and J. Ferrer, Phys. Rev. B {\bf 81},  125447 (2010).


\bibitem{Michalak2010}{\L}. Michalak, C. M. Canali, M. R. Pederson, M. Paulsson, and V. G. Benza, 
Phys. Rev. Lett. {\bf 104}, 017202 (2010). 
\bibitem{PedersonEffectEtralElectron}K. Park, M. R. Pederson, Phys. Rev. B {\bf 70}, 054414 (2004).


\bibitem{Mn12WithoutCH3} Z. Zeng, D. Guenzburger, and D. E. Ellis, Phys. Rev. B {\bf 59}, 6927 (1999).

\bibitem{Mn12WithHydrogenPederson} M. R. Pederson and S. N. Khanna, Phys. Rev. B  {\bf 60}, 9566 (1999).

\bibitem{Mn12WithHydrogen} 
M. R. Pederson, D. V. Porezag, J. Kortus, and S. N. Khanna, J. Appl. Phys.  {\bf 87}, 5487 (2000);
M. R. Pederson, N. Bernstein, and J. Kortus, Phys. Rev. Lett.  {\bf 89}, 097202 (2002);
D. W. Boukhvalov, A. I. Lichtenstein, V. V. Dobrovitski, M. I. Katsnelson, B. N. Harmon, V. V. Mazurenko, and V. I. Anisimov, Phys. Rev. B  {\bf 65}, 184435 (2002).


\bibitem{LigandEffectsHan2004}M. J. Han, T. Ozaki, and J. Yu, Phys. Rev. B {\bf 70}, 184421 (2004)


\bibitem{LDAUBoukhvalov2002} D. W. Boukhvalov, A. I. Lichtenstein, V. V. Dobrovitski, M. I. Katsnelson, B. N. Harmon, V. V. Mazurenko, and V. I. Anisimov, Phys. Rev. B {\bf 65}, 184435 (2002).
\bibitem{LDAUBoukhvalov2004} D. W. Boukhvalov, V. V. Dobrovitski, M. I. Katsnelson, A. I. Lichtenstein, B. N. Harmon, P. K\"{o}gerler,  Phys. Rev. B {\bf 70}, 054417 (2004).

\bibitem{LDAUPennino} U. del Pennino, V. De Renzi, R. Biagi, V. Corradini, L. Zobbi, A. Cornia, D. Gatteschi, F. Bondino, E. Magnano, M. Zangrando, M. Zacchigna, A. Lichtenstein, and D. W. Boukhvalov, Surf. Sci. {\bf 600}, 4185 (2006).

\bibitem{LDAUBarbour}A. Barbour, R. D. Luttrell, J. Choi, J. L. Musfeldt, D. Zipse, N. S. Dalal, D. W. Boukhvalov, V. V. Dobrovitski, M. I. Katsnelson, A. I. Lichtenstein, B. N. Harmon, and P. K\"{o}gerler, Phys. Rev. B {\bf 74}, 014411 (2006).

\bibitem{ExpTheoU}D. W. Boukhvalov, M. Al-Saqer, E. Z. Kurmaev, A. Moewes, V. R. Galakhov, L. D. Finkelstein, S. Chiuzb\u{a}ian, M. Neumann, V. V. Dobrovitski, M. I. Katsnelson, A. I. Lichtenstein, B. N. Harmon, K. Endo, J. M. North, and N. S. Dalal, Phys. Rev. B {\bf75}, 014419 (2007).


\bibitem{prelim} F. Rostamzadeh Renani and G. Kirczenow Phys. Rev {\bf B} 84, 180408(R) (2011).

\bibitem{Experiment1ElectronEppley} H. J. Eppley, H.-L. Tsai, N. de Vries, K. Folting, G. Christou, and D. N. Hendrickson, J. Am. Chem. Soc. {\bf 117}, 301 (1995).


\bibitem {huckel_off_diagonal} J. H. Ammeter, H. B. Buergi, J. C. Thibeault, and R. Hoffmann, J. Am. Chem. Soc.{\bf 100 }, 3686 (1978).

\bibitem{YAEHMOP}The numerical implementation used was the YAEHMOP package version 3.0, by G. A. Landrum and W. V. Glassey,  with the default extended H\"{u}ckel prescription for that program, given in Ref \onlinecite{huckel_off_diagonal}. YAEHMOP does not include spin-orbit coupling.

\bibitem{Datta1997}S. Datta, W. Tian, S. Hong, R. Reifenberger, J. I. Henderson, and C. P. Kubiak, Phys. Rev. Lett. {\bf 79}, 2530 (1997).

\bibitem{EmberlyKirczenow01}E. G. Emberly, G. Kirczenow, Phys. Rev. Lett.
{\bf 87}, 269701 (2001); Phys. Rev. B {\bf 64}, 235412 (2001).

\bibitem{Kushmerick02} J.G. Kushmerick, D.B. Holt, J.C. Yang, J. Naciri, M.H.
Moore, and R. Shashidhar, Phys. Rev. Lett. {\bf 89}, 086802 (2002).

\bibitem{Cardamone08}D. M. Cardamone and G. Kirczenow,  Phys. Rev. B {\bf 77}, 165403 (2008).

\bibitem{Demir2011} F. Demir and  G. Kirczenow, J. Chem. Phys. {\bf 134}, 121103 (2011).

\bibitem{PivaWolkowKirczenow05}G. Kirczenow, P. G. Piva and R. A. Wolkow, Phys.
Rev. B  {\bf 72}, 245306 (2005).

\bibitem{PivaWolkowKirczenow08} P. G. Piva, R. A. Wolkow, G. Kirczenow,  Phys.
Rev. Lett. {\bf 101}, 106801 (2008).

\bibitem{PivaWolkowKirczenow09}G. Kirczenow, P. G. Piva and R. A. Wolkow,  Phys.
Rev. B {\bf 80}, 035309(2009).

\bibitem{Buker08}J. Buker and G. Kirczenow,  Phys. Rev. B  {\bf 78}, 125107
(2008).

\bibitem{Buker05}J. Buker and G. Kirczenow,  Phys. Rev. B {\bf 72}, 205338
(2005).

\bibitem{Ihnatsenka11} S. Ihnatsenka and G. Kirczenow, Phys. Rev. B{\bf 83}, 245442 (2011).

\bibitem{EmberlyKirczenow02} E. G. Emberly and G. Kirczenow, Chem. Phys. 281, 311 (2002).

\bibitem{DalgleishKirczenow05a}H. Dalgleish and G. Kirczenow, Phys. Rev. B 72, 184407 (2005).

\bibitem{DalgleishKirczenow06a}H. Dalgleish and G. Kirczenow, Phys. Rev. B 73, 235436 (2006).


\bibitem{SpinOrbitDFTKortus} Jens Kortus , Mark R. Pederson, Tunna Baruah, N. Bernstein, C.S. Hellberg, Polyhedron {\bf 22}, 1871 (2003) 
\bibitem{SpinOrbitDFTPark} K. Park, M. R. Pederson, S. L. Richardson, N. Aliaga-Alcalde, G. Christou, Phys. Rev. B {\bf 68}, 020405 (2003).
\bibitem{SpinOrbitDFTSIESTA} L. Fernndez-Seivane, M. A. Oliveira, S. Sanvito, and J. Ferrer, J. Phys.: Condens. Matter {\bf 18}, 7999 (2006). 

%\bibitem{SpinOrbit} For example, see S. Gasiorowicz, Quantum Physics (Wiley, New York, 2003), p. 188.

\bibitem{Kittel} See C. Kittel, {\em Quantum Theory of Solids} (Wiley, New York, 1963), p. 181.

\bibitem{SpinOrbiGraphene} S. Konschuh, M. Gmitra,  J. Fabian, Phys. Rev. B {\bf 82}, 245412 (2010).

\bibitem{SOMnTheory} M. Vijayakumar, M. S. Gopinathan, J. Mol. Struct. (THEOCHEM)  {\bf 361}, 15 (996); E. Francisco and L. Pueyo, Phys. Rev. B {\bf 37}, 5278(1988)J. P. Desclaux, At. Data Nucl. Data Tables, 1{\bf 2}, 311 (1973); E. Francisco and L. Pueyo, Phys. Rev. A {\bf 36}, 1978 (1987); A. Abragam and B. Bleaney, \textit{Electron Paramagnetic Resonance of Transition Ions} (Clarendon Press, Oxford, 1970); M. Blume and R. E. Watson, Proc. R. Soc. London, Sect. A {\bf 271}, 565 (1963);  F. Herman and S. Skillman, \textit{Atomic Structure Calculations} (Prentice-Hall, Inc., Englewood Cliffs, N. J., 1963).


\bibitem{SOMnExperiment}J. Bendix, M. Brorson, C. E. Schaffer, Inorg. Chem. {\bf 32}, 2838-2849 (1993);
CRC Handbook of Chemistry and Physics (CRC Press, Boca Raton, 19871988), 68th ed.;
W. U. L. Tchang-Brillet, M. C. Artru, J. F. Wyart, Phys. Scr. {\bf 33} 390 (1986);
M. Gerloch, \textit{Orbitals, Terms and States} (Wiley, Chichester, 1986)
C. Corliss and J. Sugar, J. Phys. Chem. Ref. Data 6, 1253 (1977);
 T. A. Carlson, \textit{Photoelectron and Auger Spectroscopy} (Plenum, New York, 1975); 
 G. Mattney Cole Jr., Barry B. Garrett,  Inorg. Chem. {\bf 9}, 1898 (1970 ); 
J. Ferguson, Aust. J. Chem. {/bf 21}, 307 (1968); 
A. Mehra and P. Venkateswarlu, Phys. Rev. Lett. {\bf 19}, 145 (1967);
D. S. McClure, Solid State Phys. {\bf 9}, 399 (1959); 


\bibitem{Geometry_data_Me} D. Ruiz-Molinaa, P. Gerbiera, E. Rumbergerb, D. B. Amabilinoa, I. A. Guzeic, K. Foltingd, J. C. Huffmand, A. Rheingoldc, G. Christou, J. Veciana and D. N. Hendrickson, J. Mater. Chem. {\bf  12}, 1152 (2002).

\bibitem{SpinMn12Me}R. Sessoli, H-L. Tsai, A. R. Schake, S. Wang, J. B. Vincent, K. Folting, D. Gatteschi, G. Chistou, and D. N. Hendrickson, J. Am. Soc.{\bf 115} 1804 (1993).



\bibitem{Geometry_data_Et_and_Ph}S. M. J. Aubin, Z. Sun, H. J. Eppley, E. M. Rumberger, I. A. Guzei, K. Folting, P. K. Gantzel, A. L. Rheingold, G. Christou, and D. N. Hendrickson, Inorg.Chem. {\bf  40}, 2127 (2001).

\bibitem{MABneutralMn12Ph}K. Takeda, K. Awaga, and T. Inabe, Phys. Rev. B {\bf 57}, R11062 (1998).

\bibitem{MnAcExperimentalMAB} S. M. Oppenheimer, A. B. Sushkov, J. L. Musfeldt, R. M. Achey, and N. S. Dalal, Phys. Rev. B 65, 054419 (2002);
J. M. North, D. Zipse, N. S. Dalal, E. S. Choi, E. Jobiliong, J. S. Brooks, and D. L. Eaton, Phys. Rev. B 67, 174407 (2003).
 
 

\bibitem{DFTEnergyGap}C. Franchini, V. Bayer, R. Podloucky, J. Paier, and G. Kresse, Phys. Rev. B {\bf 72}, 045132 (2005).

\bibitem{FermiEenrgy}The Fermi energy $E_{\mbox{\scriptsize{F}}}$ of the gold contacts  has been calculated by performing extended H\"{u}ckel theory-based computations for large gold clusters.

\bibitem{Experiment1ElectronAubin} S. M. J. Aubin, Z. Sun, L. Pardi, J. Krzystek, K. Folting, L.-C. Brunel, A. L. Rheingold, G. Christou, D. N. Hendrickson. Inorg. Chem. {\bf 38}, 5329 (1999).
\bibitem{Experiment2ElectronSoler}M. Soler, W. Wernsdorfer, K. A. Abboud, J. C. Huffman, E. R. Davidson, D. N. Hendrickson, and G. Christou, J. Am. Chem. Soc. {\bf 125}, 3576 (2003).

\bibitem{MABchargedMn12Ph} K. Takeda , K. Awaga , Phys. Rev. B {\b f56} , 14560 (1997).

\bibitem{Cardamone10}D. M. Cardamone and G. Kirczenow,  Nano Lett.  {\bf 10} , 1158 (2010).

\bibitem{Kirczenow07}G. Kirczenow, Phys. Rev. B {\bf75}, 045428 (2007).

\bibitem{DalgleishKirczenow06} H. Dalgleish and G. Kirczenow, Nano Lett. {\bf 6} 1274 (2006).

\bibitem{DalgleishKirczenow05}H. Dalgleish and G. Kirczenow, Phys. Rev. B  {\bf 72}, 155429 (2005).

\bibitem{EmberlyKirczenow98} E. Emberly and G. Kirczenow, Phys. Rev. Lett. {\bf 81}, 5205-5208 (1998).

\bibitem{polyacetylene}
C. R. Fincher, Jr., M. Ozaki, M. Tanaka, D. Peebles, L. Lauchlan, A. J. Heeger, 
A. G. MacDiarmid, Phys. Rev. B 20, 1589 (1979).

\bibitem {polythiophene} K. Lee, G. K. Sotzing, Macromolecules 34, 5746 (2001).


}
\end{thebibliography}
